\documentclass[prd,twocolumn,amsmath,amssymb,floatfix,superscriptaddress,nofootinbib,preprintnumbers]{revtex4-1}
\usepackage[utf8]{inputenc}
\usepackage{amssymb}
\usepackage{amsmath}
\usepackage[toc,page]{appendix} 
\usepackage{bm}
\usepackage{color}
\usepackage{enumitem}
\usepackage{booktabs}
\usepackage{graphicx}

\usepackage[labelfont=bf]{caption}
\usepackage{subcaption}

\usepackage{pifont}

\usepackage{extarrows}

\usepackage[table, svgnames, dvipsnames]{xcolor}
\usepackage{makecell, cellspace, caption}
\definecolor{aliceblue}{rgb}{0.94, 0.97, 1.0}
\definecolor{cambridgeblue}{rgb}{0.64, 0.76, 0.68}
\definecolor{celadon}{rgb}{0.67, 0.88, 0.69}
\definecolor{classicrose}{rgb}{0.98, 0.8, 0.91}

\DeclareMathAlphabet\mathbfcal{OMS}{cmsy}{b}{n}
\definecolor{darkgreen}{RGB}{50,150,0}
\definecolor{purple}{cmyk}{0.5,0.75,0,0}
\definecolor{darkpurple}{RGB}{128,0,128}
\definecolor{ultramarine}{rgb}{0.07, 0.04, 0.56}
\definecolor{cadmiumgreen}{rgb}{0.0, 0.42, 0.24}
\definecolor{indigo(dye)}{rgb}{0.0, 0.25, 0.42}

\usepackage[linktocpage=true]{hyperref}
\hypersetup{
colorlinks=true,
citecolor=ultramarine,
linkcolor=cadmiumgreen,
urlcolor=indigo(dye),
pdfauthor={},
pdftitle={},
pdfsubject={}
}

\def\aj{{\rm AJ}}                   % Astronomical Journal
\def\mnras{{\rm MNRAS}}             % Monthly Notices of the RAS
\def\prd{{\rm Phys.~Rev.~D}}        % Physical Review D
\def\mnras{Mon.\ Not.\ R.\ Astron.\ Soc.\ }
\def\jcap{{\rm JCAP}}

\begin{document}

\title{Flow-based likelihoods for non-Gaussian inference}

\author{Ana Diaz Rivero}
\email{adiazrivero@g.harvard.edu}
\affiliation{Department of Physics, Harvard University, Cambridge, MA 02138, USA}

\author{Cora Dvorkin}
\email{cdvorkin@g.harvard.edu}
\affiliation{Department of Physics, Harvard University, Cambridge, MA 02138, USA}

\begin{abstract}

\noindent We investigate the use of data-driven likelihoods to bypass a key assumption made in many scientific analyses, which is that the true likelihood of the data is Gaussian. In particular, we suggest using the optimization targets of flow-based generative models, a class of models that can capture complex distributions by transforming a simple base distribution through layers of nonlinearities. We call these flow-based likelihoods (FBL). We analyze the accuracy and precision of the reconstructed likelihoods on mock Gaussian data, and show that simply gauging the quality of samples drawn from the trained model is not a sufficient indicator that the true likelihood has been learned. We nevertheless demonstrate that the likelihood can be reconstructed to a precision equal to that of sampling error due to a finite sample size. We then apply FBLs to mock weak lensing convergence power spectra, a cosmological observable that is significantly non-Gaussian (NG). We find that the FBL captures the NG signatures in the data extremely well, while other commonly used data-driven likelihoods, such as Gaussian mixture models and independent component analysis, fail to do so. This suggests that works that have found small posterior shifts in NG data with data-driven likelihoods such as these could be underestimating the impact of non-Gaussianity in parameter constraints. By introducing a suite of tests that can capture different levels of NG in the data, we show that the success or failure of traditional data-driven likelihoods can be tied back to the structure of the NG in the data. Unlike other methods, the flexibility of the FBL makes it successful at tackling different types of NG simultaneously. Because of this, and consequently their likely applicability across datasets and domains, we encourage their use for inference when sufficient mock data are available for training.

\end{abstract}

\maketitle

\section{Introduction}\label{sec:introduction}

There are three key ingredients when doing inference: data, a model, and a likelihood function (in the case of Bayesian inference, there is also a prior). The quality of the inferred parameters of course hinges on the quality of the data, how closely the model approximates the process that gave rise to the data, and how accurately the likelihood maps the probability of observing the data given the model. Throughout this work, we are going to be concerned with the last of these. 

In data analysis across many disciplines the likelihood used for inference is often assumed to be Gaussian. Gaussian likelihoods are attractive and widespread because they are well understood and inference boils down to obtaining a covariance matrix for the data. However, in reality, it is not generally true that the underlying likelihood of a dataset is Gaussian. 

There are several general points one can keep in mind when considering the applicability of a Gaussian likelihood function. One is the central limit theorem (CLT), which states that the difference between a sample mean and a true mean, normalized by the standard error, tends toward the standard normal distribution as the number of samples tends to infinity. This is generally invoked in favor of using Gaussian likelihoods, even when datasets are known to be non-Gaussian. For example, for data in Fourier space (such as power spectra), many modes contribute at high wave numbers and thus the CLT approximately applies. Conversely, at low wave numbers few modes contribute and the CLT is not applicable. Another important point is that, in works where the covariance matrix of the data is an estimated quantity, meaning it is not known \textit{a priori}, it becomes a stochastic object with some uncertainty. Thus, to obtain the likelihood of the data given the estimated covariance matrix, the likelihood function has to be marginalized over the true (unknown) covariance, conditioned on the estimated one.
If the original likelihood is assumed to be Gaussian, this marginalization step in fact leads to a multivariate $t$-distribution \cite{Sellentin_tstudent}. Finally, it is also important to consider that systematic effects might introduce correlations in the data that are not Gaussian. 

There are also field-specific factors that can inform the choice of likelihood function. Although the method we develop in this work is  applicable for inference with any dataset, here we apply it to a cosmological one. There are several puzzling tensions in different cosmological datasets that have so far stood the test of time \cite{S8_tension,Tensions,features} that would be interesting to reconsider under a new, more accurate likelihood. For cosmological data, we can use knowledge about the physical processes that give rise to an observable to understand its Gaussianity. For instance, while some cosmological fields such as the cosmic microwave background (CMB) \textit{are} essentially Gaussian, those that follow from nonlinear gravitational collapse $-$ such as distributions of galaxies and cosmic shear $-$ are highly non-Gaussian. It is therefore likely that the underlying likelihoods of such fields are themselves non-Gaussian: a nonlinear function of a Gaussian random variable is not Gaussian distributed. 

Recently, several works have studied the impact of using Gaussian likelihoods to infer parameters from non-Gaussian cosmological data, showing that it can bias the posteriors of inferred cosmological parameters and underestimate uncertainties \cite{Hahn:2018zja, Sellentin:2017fbg,Sellentin_tstudent,sellentin_starck}. 
The reason behind this is quite intuitive: the use of Gaussian likelihoods to model non-Gaussian processes becomes a source of systematic error. 

One promising approach to avoid relying on a Gaussian likelihood is to use what are called \textit{data-driven likelihoods} (DDL): likelihoods that are learned directly from the data.\footnote{We detail two points with respect to this terminology. We use the term DDL to refer to likelihoods that are more flexible than a multivariate Gaussian (or other ubiquitous, simple parametric likelihoods) and do not necessarily fix the functional form at the outset. Note that, while for a MVN likelihood the covariance matrix can be estimated from data too, the functional form of the likelihood is otherwise fixed. The second point is what is meant by 
``data". We only have one universe, so in cosmology we often have to rely on mock (simulated) realizations of the universe to assess the fit of a model to some data. Because the method presented in this paper is applied in the context of cosmology, we will technically be using \textit{mock data}-driven likelihoods. But because our method is equally applicable in any field, one could imagine disciplines where experiments are repeatable and thus the likelihoods are indeed learned from real data.}
Nevertheless, there is not yet a unified framework to deal with non-Gaussian likelihoods. Instead, data-driven approaches rely heavily on trial and error, and tend to not generalize well in different contexts \cite{Hahn:2018zja}. In fact, we will show that depending on the type of non-Gaussianity in the data some DDL methods might not be appropriate and thus offer little improvement upon a multivariate normal (MVN) likelihood. 

Here, we use a \textit{flow-based generative model} \cite{ffjord} to achieve this task. Generative models are models from which new samples can be drawn when trained, by for example learning the underlying true likelihood that gave rise to the data. Flow-based models, in particular, consist of applying a series of transformations (a flow) to a simple probability density that is easy to sample from (e.g., a univariate Gaussian) to transform it into a (theoretically) arbitrarily complex probability density function \cite{normalizing_flows}. If the transformations are known, an exact form for the resulting likelihood can be obtained and sampled from. By parametrizing the transformations with neural networks, these models can be made very expressive (in this context, 
``expressive" refers to a model that can capture complex features in the data accurately). Flow-based models have been used to enrich variational posteriors or enhance other generative models (e.g., \cite{normalizing_flows, nf_reion, bohm2020probabilistic, brehmer2020flows, dai2020sliced}) and for density estimation, to learn complicated probability distributions (e.g., governing images of human faces) \cite{papamakarios2018masked,nice,RealNVP,glow,gbs,durkan2019neural}. The driving interest behind these models has been their capacity to learn complicated probability distributions, but understanding how faithfully they reconstruct the true data distribution is an open area of research \cite{nalisnick2019deep,bohm2019uncertainty}. 

We seek to understand the quality of the learned likelihood function with the goal of establishing \textit{flow-based likelihoods} (FBLs) as an extremely flexible and adaptable method to obtain data-driven likelihoods that are more accurate than other methods to deal with known (or suspected) non-Gaussianities in datasets. FBLs have appeared in the machine learning and cosmology/astrophysics literature in other settings, such as in likelihood-free inference (e.g., DELFI)  \cite{pmlr-v89-papamakarios19a,delfi1,delfi2,durkan2018sequential,pmlr-v96-lueckmann19a} and simulation-based inference \cite{Cranmer201912789,Tejero-Cantero2020,Wong_2020,cranmer2019modeling}. We demonstrate that FBLs can match the underlying likelihood of the data very well, even in high-dimensional spaces, by comparing the non-Gaussian features of the generated samples and the data. Furthermore, we show that this is not trivial: under certain conditions, the samples generated can be indistinguishable from the original data and yet the likelihood can be significantly biased and imprecise. We apply the FBL to mock weak lensing power spectra, which we show have strong non-Gaussianities. This observable is representative of many observables in cosmology for which, given a set of cosmological model parameters, we can generate sophisticated mock data with forward simulations (including instrumental effects, selection biases, and other systematics) but we cannot write down a tractable likelihood. Along the way we create a thorough pipeline, incorporating some elements from previous works \cite{Hahn:2018zja,Sellentin:2017fbg}, which can be applied to any dataset to analyze the presence (or absence) of non-Gaussianities. With this pipeline we are also able to gain some insight into why some data-driven methods work for some datasets and not for others. Seeing that FBLs are capable of overcoming barriers that other data-driven methods can face depending on the type of non-Gaussianity in the data further emphasizes the advantage of using FBLs for inference. \textit{We emphasize that no aspect of the NG pipeline nor FBLs are specific to cosmology and therefore these methods can be applied in any context.}

In Section \ref{sec:ddls} we briefly present two data-driven likelihood methods, Gaussian mixture models (GMMs) and independent component analysis (ICA), which we use to compare to our FBL. We also outline the basic principles behind normalizing flows and describe FFJORD \cite{ffjord}, a flow-based generative model that uses ordinary differential equations to evolve the initial probability density. In Section \ref{sec:nongauss} we present the exhaustive tests we carry out to characterize non-Gaussianity in the data. In Section \ref{sec:wl_dset} we introduce the observable to which we apply our flow-based likelihood (FBL), the weak lensing convergence power spectrum, and finally we show our results applying the FBL to this data in Section \ref{sec:apply_flow}. We discuss our results and conclude in Sections \ref{sec:discussion} and \ref{sec:conclusion}. All the code used in this work will be made available upon publication.

Throughout the remainder of the paper, we will use the terms Gaussian, multivariate Gaussian, and multivariate normal interchangeably. Furthermore, we will abbreviate ``non-Gaussian" (and its derivatives) as NG when convenient. 

\section{Data-Driven Likelihoods}
\label{sec:ddls}

A probability density function (PDF) can be estimated by drawing sufficient samples from it. This is the key idea behind data-driven likelihoods. With access to sufficient (mock) data, it is not necessary to impose a restrictive functional form for the likelihood \textit{a priori}; instead, we can think of independent (mock) catalogs as independent draws from the underlying true likelihood function, and can use this fact to estimate the likelihood directly from the samples. The benefit of this method is that, if an appropriate method for density estimation is used, the reconstructed likelihood can take into account any non-Gaussian features in the data.\footnote{We are using the terms PDF and likelihood interchangeably, although they are different objects. We explain this in Appendix \ref{app:likelihood_vs_sampling_density}.} The drawback is that there is no guarantee that these methods converge on the true likelihood, and extensive validation is required. Our validation procedure is addressed in Sections \ref{sec:nongauss} and \ref{sec:apply_flow}.

We refer to likelihoods learned from the data as data-driven likelihoods (DDL). In this work, we implement three different DDLs. The first two, used as our baseline, were chosen to serve as a direct comparison of the work in Ref. \cite{Hahn:2018zja}, which studied NG in large-scale structure data. These are Gaussian mixture models and independent component analysis. We briefly summarize key aspects of each of these in the section below, but due to their ubiquity in the literature we provide references for more detailed explanations. The third method is what we refer to as flow-based likelihoods (FBL): we propose using the likelihood learned by flow-based generative models as a DDL. We discuss this in detail below. 

\subsection{Gaussian mixture model}

As its name indicates, a Gaussian mixture model (GMM) is simply a convex combination (mixture) of multiple Gaussians with unknown means and covariances, each with a weight that determines their contribution to the PDF. In a multidimensional setting, the final PDF for a vector $\mathbf{x}$ in a GMM with $K$ components can be written as

\begin{equation}
    \hat{p}_{\rm GMM}(\mathbf{x}) = \sum_{i=1}^K \phi_i \mathcal{N}(\mathbf{x}|\mu_i, \Sigma_i),
\end{equation}

\noindent where $\mu_i$, $\Sigma_i$ and $\phi_i$ are the mean, covariance, and weight of the $i$th Gaussian in the mixture, respectively, and $\mathcal{N}$ is the multivariate normal PDF. The weights are normalized such that $\sum_{i=1}^K \phi_i = 1$. The number of parameters in these models is given by 

\begin{equation}
    K \left(d + \frac{1}{2} d (d+1) \right) + K,
\end{equation}

\noindent where $d$ is the dimension of the data vector $\mathbf{x}$. This is because for each of the $K$ components we have to learn a $d$-dimensional mean vector and a covariance matrix with $\frac{1}{2} d (d+1)$ degrees of freedom (since it is positive semidefinite)\footnote{The number of parameters can also be decreased by putting additional constraints on the covariance matrix, such as having different components share a covariance matrix or forcing it to be diagonal, but we will use full covariances throughout this paper.}. We also have to learn $K$ weights, one for each component. Notice that for $K=1$ we recover a standard MVN likelihood. 

We use the \texttt{scikit-learn} \cite{scikit-learn} implementation of GMMs, which uses expectation maximization \cite{em} to estimate the model parameters, and the Bayesian information criterion (BIC) to decide how many components to include in the mixture. This method of model selection considers the maximum likelihood of a model while penalizing model complexity. We refer the reader to Ref. \cite{bishop} for a thorough explanation of GMMs and the expectation maximization procedure.

\subsection{Independent component analysis}

Independent component analysis (ICA) is typically used to isolate linear mixtures of independent sources. For some observed data vector $\mathbf{x}$, this amounts to the assumption

\begin{equation}
    \mathbf{x} = \mathbf{A}\mathbf{s},
\end{equation}

\noindent where $\mathbf{A}$ is an unknown matrix, called the mixing matrix, that mixes the sources $\mathbf{s}$. The goal of ICA is to solve for $\mathbf{A}$, which is done by actually solving for its inverse, $\mathbf{A^{-1}} = \mathbf{W}$,

\begin{equation}
    \hat{\mathbf{s}} = \mathbf{W}\mathbf{x},
\end{equation}

\noindent such that $\hat{\mathbf{s}} \approx \mathbf{s}$. 

ICA solves for $\mathbf{W}$ by breaking it up into three different linear operations. The first one is decorrelating the data (i.e. projecting the data onto the principal components), an operation usually referred to as principal components analysis. The data is then normalized, and finally a rotation matrix is solved for such that the statistical independence of the sources is maximized. We refer the readers to Ref. \cite{ica} for details on how this maximization is carried out. We use the \texttt{scikit-learn} implementation of the ICA algorithm. 

For our purposes, what matters is that ICA provides a way of turning the $d$-dimensional likelihood for $\mathbf{x}$ into $d$ one-dimensional likelihoods by converting $\mathbf{x}$ into $d$ independent components\footnote{Recall that statistical independence requires that neither second-order nor higher-order correlations exist.}:

\begin{equation}
    \hat{\mathbf{s}} \equiv \mathbf{x}_{\rm ICA} = \mathbf{W}\mathbf{x} = \{\mathbf{x}_{\rm 1,ICA},...,\mathbf{x}_{N,\rm{ICA}} \},
\end{equation}

\noindent where $N$ is the number of ICA components. In this work, we set $N=d$, although in principle ICA can also be used for dimensionality reduction. 

The one-dimensional probability density of each component, $\hat{p}_n$, can then be estimated using a kernel density estimator (KDE). We use a Gaussian kernel, with a standard deviation\footnote{More generally, kernels are controlled by a parameter called the bandwidth. It decides how much to smooth each data point, and therefore controls the bias-variance trade-off. For the case of a Gaussian kernel, the bandwidth is the same as the standard deviation.} given by Scott's bandwidth \cite{scott}. 
Finally, the likelihood for $\mathbf{x}$ is a factorial distribution, a product of the $N$ independent PDFs:

\begin{equation}
    \hat{p}_{\rm ICA}(\mathbf{x}) = \prod_{n=1}^{N} \hat{p}_{n}(\mathbf{x}).
\end{equation}

\subsection{Flow-based likelihood}\label{sec:fbl}

\subsubsection{Flow-based generative models}

\begin{figure*}
    \centering
    \includegraphics[width=0.85\textwidth]{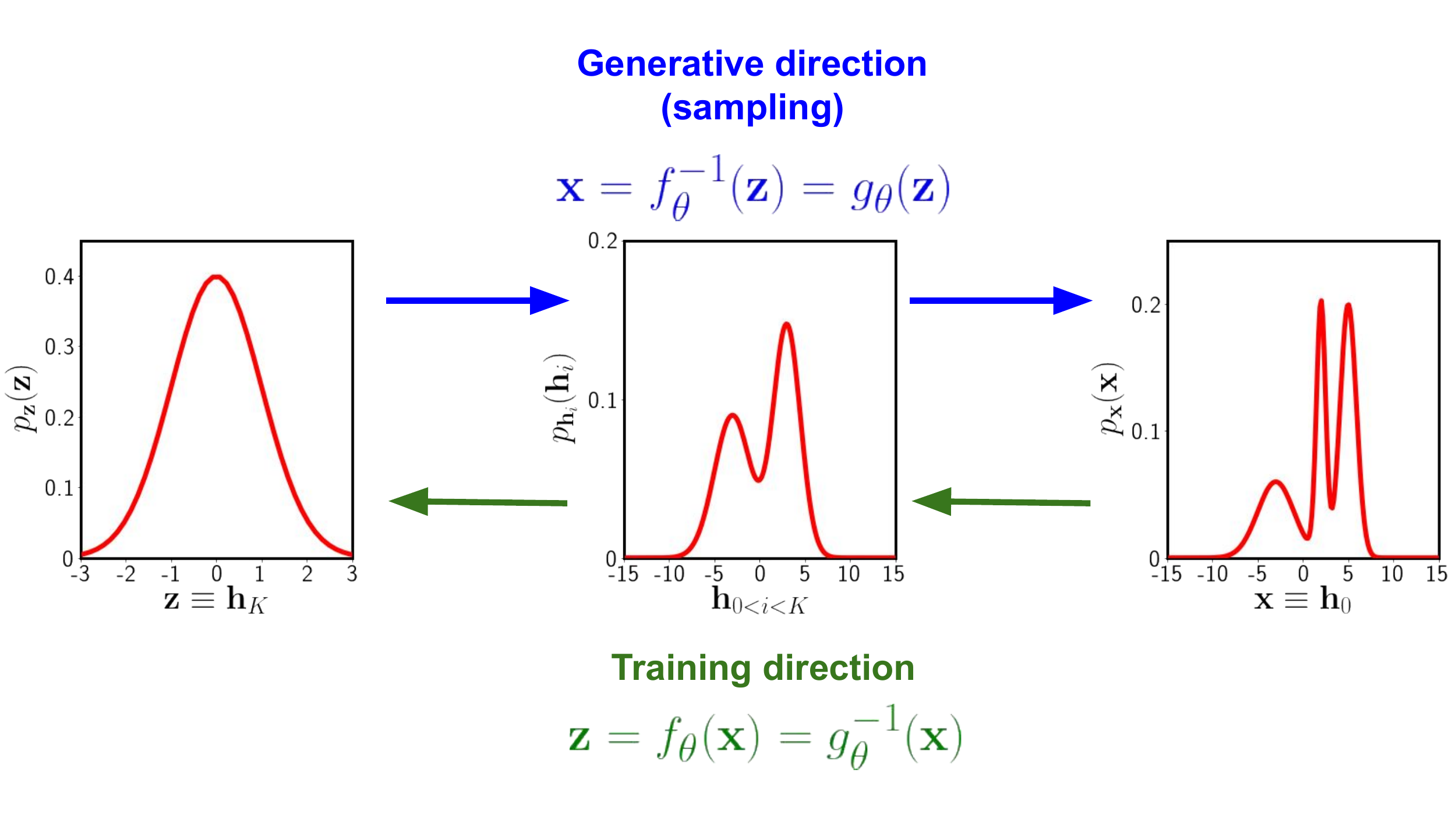}
    \caption{\footnotesize{Schematic depiction of a flow-based generative model in one dimension. The left-hand panel shows a standard normal, which is chosen as the base distribution $p_{\mathbf{z}}(\mathbf{z} \equiv \mathbf{h}_K)$, following the notation in Eq. \eqref{eq:flow}. The right-hand panel shows the target distribution, i.e. the distribution of the data $p_{\mathbf{x}}(\mathbf{x} \equiv \mathbf{h}_0)$, which is visibly more complex than the prior. The middle panel shows the distribution at an intermediate point in the flow, $\mathbf{h}_{0 < i < K}$. The green arrows reflect the direction in which training takes place: the model is fed the data $\mathbf{x}$, which then undergoes the flow into the latent space. The blue arrows reflect the generative, or sampling, direction, whereby a sample is drawn from the prior distribution and undergoes the inverse flow to generate a data sample.}} \label{fig:flow}
\end{figure*}

For a given data vector $\mathbf{x}$, the goal of generative models is to learn the distribution that $\mathbf{x}$ is drawn from: $\mathbf{x} \sim p^*(\mathbf{x})$, where the asterisk denotes the true underlying distribution. Once an estimate of $p^*$ is obtained, new samples of $\mathbf{x}$ can be generated, thus the name of this class of models. 

In flow-based methods, the generative process starts by drawing a sample $\mathbf{z}$ from a (simple) probability distribution that has a tractable PDF and is easy to sample from (such as a univariate Gaussian), 
\begin{equation}
   \mathbf{z} \sim p_{\mathbf{z}}(\mathbf{z}).
\end{equation}

\noindent This prior distribution then undergoes a ``flow", which means that it is transformed repeatedly while conserving its probability. The optimization process relies on finding a series of transformations such that the resulting distribution approaches $p^*(\mathbf{x})$. 

Consider a series of transformations that are bijective, so that the relationship between $\mathbf{z}$ and $\mathbf{x}$ can be summarized as

\begin{equation}\label{eq:flow}
\mathbf{x} \equiv \mathbf{h}_0 \xleftrightarrow[g_{K}]{f_1} \mathbf{h_1} \xleftrightarrow[g_{\rm K-1}]{f_2} ... \xleftrightarrow[g_{2}]{f_{K-1}} \mathbf{h_{K-1}} \xleftrightarrow[g_1]{\rm f_{K}} \mathbf{h_{K}} \equiv \mathbf{z},
\end{equation}

\noindent such that $g = g_1 \circ g_2 \circ ... g_{K}$, $f = f_1 \circ f_2 \circ ... f_{K}$ and $\mathbf{z} = f_{\theta}(\mathbf{x}) = g_{\theta}^{-1}(\mathbf{x})$. If the transformation $f_{\theta}(\mathbf{x})$ is learned from the data, then the invertibility criterion ensures that, after drawing a sample of $\mathbf{z}$, $f_{\theta}$ can be inverted to generate new samples of $\mathbf{x}$. Furthermore, notice that enforcing the transformations to be bijective enforces volume preservation (and unlike other generative models such as variational autoencoders \cite{vae}, the dimension of the latent space and the data is the same). In theory, if the transformations are expressive enough, $p_{\mathbf{z}}$ can be turned into an arbitrarily complex distribution. To this end, the transformations tend to involve (invertible) neural networks. Figure \ref{fig:flow} depicts how flow-based generative models work in a simplified one-dimensional setting.

Ultimately, Eq. \eqref{eq:flow} is simply a concatenated sequence of changes of variables, so the log-density of the final distribution can be written as a sum of the log-PDF of the original distribution plus the sum of the log-determinant of the Jacobian matrix of each transformation:

\begin{align}
\log p_{\mathbf{x}}(\mathbf{x}) &= \log p_{\mathbf{z}}(\mathbf{z}) + \log \bigg \rvert \det\left(\frac{d\mathbf{z}}{d\mathbf{x}} \right)\bigg\rvert \\
&= \log p_{\mathbf{z}}(\mathbf{z}) + \sum_{i=1}^{K} \log \bigg \rvert \det\left(\frac{d\mathbf{h}_i}{d\mathbf{h}_{i-1}} \right)\bigg\rvert.
\end{align}

\noindent The loss function used for training is then simply the negative log-likelihood over the entire training set $\mathcal{D}$:
\begin{equation}
\mathcal{L} = -\frac{1}{|\mathcal{D}|} \sum_{\mathbf{x} \in \mathcal{D}} \log p(\mathbf{x}; \mathbf{w}) ,
\end{equation}
and the model can be trained using stochastic gradient descent to find the optimal network parameters $\mathbf{w}$. 

There are two main catches to flow-based models: one, the transformations must be easily invertible, and two, they must have an easy-to-compute Jacobian determinant, since computing the determinant of a matrix scales as $n^3$ for an $n \times n$ matrix. 

Models in the literature have tackled these issues in different ways: the form of the transformation can be restricted such that determinant identities can be exploited \cite{normalizing_flows}; the models can be made autoregressive, such that the Jacobian is lower triangular \cite{iaf}; or dimensions can be split up and affine transformations used such that the Jacobian is easy to compute \cite{nice,rnvp,glow}. All these methods have their own advantages and drawbacks, but they all have one feature in common: they sacrifice model expressivity to improve the speed of the Jacobian determinant computation. 

\subsubsection{FFJORD: Free-form Jacobian of reversible dynamics}

We use FFJORD \cite{ffjord}\footnote{https://github.com/rtqichen/ffjord}, a flow model that replaces the transformation function with an integral of continuous-time dynamics, giving rise to continuous normalizing flows (CNFs) \cite{chen}. We opted for this model because it is very expressive while remaining quick at both density estimation and sampling, which is crucial for the procedure we will carry out to assess the likelihood non-Gaussianity.
We summarize key details of FFJORD in this section but refer the reader to Refs. \cite{chen,ffjord} for additional details on the derivation and optimization procedure.

For FFJORD, the transformation from prior to data is seen as an evolution in time. Let us define the observable data as $\mathbf{z}(t_1)$ and the original sample drawn from the prior as $\mathbf{z}_0 = \mathbf{z}(t_0)$. If we then define the time evolution of $\mathbf{z}$ as an ordinary differential equation (ODE)
\begin{equation}
    \frac{\partial \mathbf{z}(t) }{\partial t} = f(\mathbf{z}(t),t;\theta),
\end{equation}

\noindent (where $f$ can be a neural network and $\theta$ its parameters) we can obtain $\mathbf{z}(t_1)$ by solving the ODE subject to the initial condition $\mathbf{z}_0 = \mathbf{z}(t_0)$. The change in log-density is given by the instantaneous change of variables formula:

\begin{equation}
    \frac{\partial \text{log} p(\mathbf{z}(t))}{\partial t} = - \text{Tr} \left( \frac{\partial f}{\partial \mathbf{z}(t)} \right),
\end{equation}

\noindent and thus the total change in log-density can be obtained by integrating across time:

\begin{equation}\label{eq:ffjord_logpx}
    \log p(\mathbf{z}(t_1)) = \text{log} p(\mathbf{z}(t_0)) - \int_{t_0}^{t_1} \text{Tr} \left( \frac{\partial f}{\partial \mathbf{z}(t)} \right) dt.
\end{equation}

Optimizing Eq. \eqref{eq:ffjord_logpx} is nontrivial, and requires a continuous-time analog to backpropagation. It can nevertheless be combined with gradient-based optimization methods to fit the parameters $\theta$. In general, computing the trace of this transformation scales as $\mathcal{O}(n^2)$, but by using an unbiased stochastic estimator of Eq. \eqref{eq:ffjord_logpx} Ref. \cite{ffjord} decreases \footnote{It is shown that the variance of the log-likelihood induced by the estimator is less than $10^{-4}$.} the complexity further to $\mathcal{O}(n)$. This makes FFJORD scalable without having to constrain the Jacobian, yielding a very expressive model.

A key point we want to make is that the focus in works that employ these model is generally on their capacity as generators, or their abstract capacity to increase the complexity of distributions (e.g., as variational posteriors). The log-likelihood is used as the training objective, and the model quality is for example judged by the quality of the samples produced (often in a qualitative fashion). Here, together with quantitatively discussing sample quality, we zero-in on the quality of the likelihood itself (showing that sample likelihood is not necessarily indicative of likelihood quality, Appendix \ref{app:toy_problems}), with the goal of establishing FBLs as very accurate and versatile DDLs that can themselves be used for inference. 

\section{Measuring Non-Gaussianity} \label{sec:nongauss}

Quantifying the non-Gaussianity of a dataset is the first step in understanding the applicability or shortcoming of applying a Gaussian likelihood for inference. We propose carrying out three different tests to quantify the NG of a given dataset, which rely on diagnosing deviations from the null hypothesis that the underlying likelihood is a MVN, and act at different ``resolutions": 
\begin{enumerate}
    \item $t$-statistic of skewness and excess kurtosis of each bin, which quantifies the NG of bins independently;
    \item transcovariance matrix \cite{Sellentin:2017fbg}, which considers the NG of all pairs of bins;
    \item the Kullback-Leibler (KL) divergence \cite{kullback1951} of all the data with respect to a MVN distribution.
\end{enumerate}

Notice that the latter two tests are sensitive to all higher-order correlations in the data. The different scopes of these three tests culminate in a very holistic indicator of NG when combined. As we will show in subsequent sections, the pipeline's tiered scope is capable of shedding light into why some DDLs work in some settings and not others, with the important consequence that methods that have applied DDLs in the past may have failed to capture non-Gaussianities adequately.

\subsection{t-statistic of skewness and excess kurtosis} 

By having many mocks of a given observable, we can obtain an estimate of the distribution for each bin (e.g., the power spectrum at a specific multipole number). We then calculate the $t$-statistic of skewness and excess kurtosis of each to diagnose a deviation away from Gaussianity (under the Gaussian assumption, the null hypothesis is zero skewness and excess kurtosis). Henceforth, for conciseness we refer to the excess kurtosis simply as kurtosis.

Recall that the $t$-statistic is basically a measure of how many standard deviations away from the null hypothesis a given measure is. For a given parameter $\beta$,
\begin{equation}
    t = \frac{\hat{\beta} - \beta_{\rm null} }{\operatorname{SE}(\hat{\beta})},
\end{equation}
where $\hat{\beta}$ is the estimated value of $\beta$, $\beta_{\rm null}$ is the value of $\beta$ under the null hypothesis, and $\operatorname{SE}(\hat{\beta})$ is the standard error of $\hat{\beta}$. $t$ is thus a dimensionless quantity that measures the deviation of the estimated parameter away from the null hypothesis in units of the standard error. 

For our purposes, $\beta$ is going to be either the skewness or the kurtosis of a sample, and $\beta_{\rm null} = 0$ under the null hypothesis that the samples are drawn from a Gaussian distribution. The variance of the skewness of a random sample of size $n$ from a normal distribution is

\begin{equation}
\operatorname {\widehat{Var}}_{\rm skew} ={\frac {6n(n-1)}{(n-2)(n+1)(n+3)}},
\end{equation}

\noindent and the variance of the kurtosis

\begin{equation}
{\displaystyle \operatorname {\widehat{Var}}_{\rm kurt} = {\frac {24n(n-1)^{2}}{(n-3)(n-2)(n+3)(n+5)}}}.
\end{equation}

\noindent Taking $\operatorname{SE} = \sqrt{\widehat{\operatorname{Var}}}$, we can thus obtain the $t$-statistic for the skewness and kurtosis of each bin. One thing to notice is that the $t$-statistic is an extensive quantity: it depends on the number of data points in a bin (or, equivalently, the number of mocks). 

\subsection{Pairwise non-Gaussianity of data points} \label{sec:sellentin}

Following Ref. \cite{Sellentin:2017fbg}, we use the basic observation that 
a sum of two independent Gaussian random variables should itself be a Gaussian random variable $-$ if $x_i \sim \mathcal{N}(\mu_i,\sigma_i^2)$ and $x_j \sim \mathcal{N}(\mu_j,\sigma_j^2)$, then $x_i + x_j \sim \mathcal{N}(\mu_i + \mu_j,\sigma_i^2 + \sigma_j^2)$, where $\mathcal{N}(\mu,\sigma^2)$ is the normal distribution with mean $\mu$ and standard deviation $\sigma$\footnote{Ref. \cite{Sellentin:2017fbg} also tests two other conditions, regarding the product and quotient of normal random variables. The expected distributions are a superposition of $\chi
^2$ random variables and the Cauchy distribution, respectively. These are both very sharply peaked distributions and we found that estimating their density with a limited number of mock samples was unreliable, so we only show results for the sum.} $-$ to test the pairwise non-Gaussianity of data bins.

To perform this test, Ref. \cite{Sellentin:2017fbg} considers the pairwise sum of all the bins in a given observable. Consider an ensemble of $N$ realizations of a $d$-dimensional observable $\mathbf{x}$. Denoting the $d$ elements of the $i$th data vector as $x_i^u$, where $i \in [1,N]$ and $u \in [1,d]$, for each pair $x_i^u,x_i^v$ with $u \neq v$ it is straightforward to obtain the sum
\begin{equation}
    s_i^{u,v} = x_i^u + x_i^v.
\end{equation}
For each pair of bins $(u,v)$, there are $N$ samples of the distribution $s_i^{u,v}$.

The $N$ samples are grouped into $b$ bins of a histogram $\mathcal{H}_b$. Under the Gaussian assumption, if $N\rightarrow \infty $ and $b \rightarrow \infty$ the histogram will tend to a Gaussian distribution. The deviation from the estimate of the density of $s_i^{u,v}$ and the expected normal distribution can be calculated using the mean integrated square error (MISE):
\begin{equation}\label{eq:mise_hist}
    \frac{1}{b} \sum_{a=1}^b [\mathcal{H}_a(s_i^{u,v}) - \mathcal{N}]^2 \equiv S^+_{u,v}.
\end{equation}

Each pair of bins can be pre-processed by mean-subtracting and whitening the $2 \times 2$ covariance matrix, to destroy all the Gaussian (second-order) correlations in the data, such that any remaining correlations are necessarily of non-Gaussian origin (in practice, this means diagonalizing the covariance matrix; we also normalize the variance of each dimension such that the final covariance matrix is the identity matrix). After these two steps each bin should be a draw from a standard univariate normal $\mathcal{N}(0,1)$ if they were originally truly Gaussian, and consequently $s_i^{u,v} \sim \mathcal{N}(0,2)$.

By finding the MISE for each pair of points, Ref. \cite{Sellentin:2017fbg} builds a \textit{transcovariance} matrix $S^+$: while covariance matrices measure Gaussian correlations between pairs of parameters/bins, the transcovariance matrix measures non-Gaussian correlations. The total contamination of the $u$th data point is then simply the sum over a column of the matrix:

\begin{equation} \label{eq:eps}
    \epsilon_u^{+} = \sum_{v \neq u} S_{u,v}^+.
\end{equation}

We carry out this procedure with one importance difference: instead of using a histogram of data points, we use a kernel density estimator instead. We prefer this methodology because while the density estimate of a histogram is strongly dependent on the number of bins, and thus applying Eq. \eqref{eq:mise_hist} requires finding the value of $b$ that minimizes $S_{u,v}^+$, the KDE is insensitive to this (although it does have other tuning parameters, as discussed in Section \ref{sec:ddls}). We therefore modify Eq. \eqref{eq:mise_hist}:

\begin{equation}\label{eq:mise}
    \frac{1}{b} \sum_{a=1}^b [\mathcal{K}_a(s_i^{u,v}) - \mathcal{N}(0,2)]^2 \equiv S^+_{u,v},
\end{equation}

\noindent where $\mathcal{K}(\cdot)$ denotes the KDE, and here $b$ is simply the number of discrete values at which we estimate the KDE and the normal distribution. We use a Gaussian kernel with standard deviation given by Scott's bandwidth. Just like for the NG test above, we note that $\epsilon^+$ is extensive, since it relies on summing over the columns of a matrix whose dimension depends on the number of data bins.

\subsection{Nonparametric Kullback-Leibler (KL) divergence} 

Following Ref. \cite{Hahn:2018zja}, we use a nonparametric estimator of the KL divergence to quantify the non-Gaussianity in a dataset. The KL divergence is a well-known measure of the (dis)similarity between two PDFs $p$ and $q$:

\begin{equation}
    D_{n,m}(p || q) = \int p(\mathbf{x}) \log\frac{p(\mathbf{x})}{q(\mathbf{x})} d\mathbf{x}.
\end{equation}

For cases in which $p$ and $q$ are unknown and we instead just have ensembles of draws from unknown distributions, Ref. \cite{kl} derived an unbiased estimator of the KL divergence that essentially relies on estimating the probability density using $k$-nearest neighbors ($k$NN). Consider two densities $p$ and $q$, defined on $\mathcal{R}^d$, and independent and identically distributed (i.i.d.) $d$-dimensional samples $\{X_1,X_2,...X_n\}$ and  $\{Y_1,Y_2,...Y_m\}$ drawn from each, respectively. Letting $\rho_k(i)$ be the Euclidean distance between $X_i$ and its $k$NN in $\{X_j\}_{j\neq i}$, and $\nu_k(i)$ the distance from $X_i$ to its $k$NN in $\{Y_j\}$, the KL divergence estimator can be
written as:

\begin{equation}\label{eq:kl}
    \hat{D}_{n,m}(p || q) = \frac{d}{n} \sum_{i=1}^n \log \frac{\nu_k(i)}{\rho_k(i)} + \log\frac{m}{n-1}.
\end{equation}

Ref. \cite{Hahn:2018zja} proposed estimating non-Gaussianity in an ensemble of mocks by comparing the nonparametric KL divergence between samples of a mock observable and samples drawn from a MVN with mean and covariance taken from the mocks. If the true likelihood that gave rise to the data were Gaussian, then the KL divergence between these two sets of samples would vanish, while deviations away from zero would indicate the presence of NG in the data.

Despite the theoretical appeal of this test, we find that for the number of data samples and bins we use, it is not quite as robust as the other two. This is due to the curse of dimensionality: $k$NN-based algorithms struggle in high-dimensional spaces due to the fast increase of volume with increasing dimensions, which makes the data sparse.

In Appendix \ref{app:kldiv} we show the variability of the KL divergence estimate when comparing two MVN distributions for different random seeds. We find that two sample distributions drawn from the same likelihood can have as little as $\sim 20\%$ overlap between them. So when showing results for this test, we will keep this lower bound in mind to judge whether a deviation can be due to NG or simply due to random error.Note that, despite these limitations, we include this test because we find that for data with significant NG (like the one used in this paper) the KL divergence estimate is much larger than the random scatter.

\begin{figure*}
    \centering
    \includegraphics[width=\textwidth]{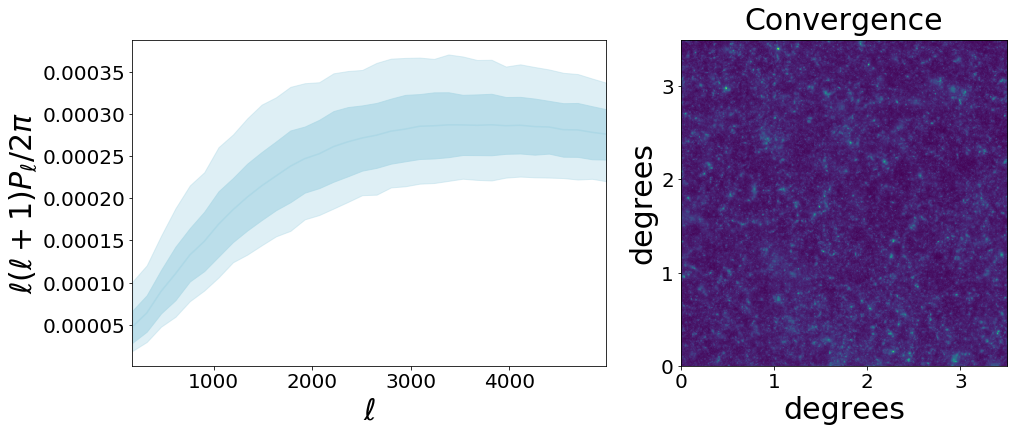}
    \caption{\footnotesize{\textit{Left}: mock convergence power spectra as a function of multipole number $\ell$. \textit{Right}: mock convergence map with a sky coverage of $3.5 \times 3.5$ deg. and $1024 \times 1024$ pixels. }} \label{fig:mock_data}
\end{figure*}

\section{Mock Weak Lensing Convergence Power Spectra}\label{sec:wl_dset}

\subsection{Background}

In the standard cosmological picture, infinitesimal initial fluctuations in the matter density field evolved through gravitational collapse in a highly nonlinear fashion to yield all the structure we have in the universe today. Mapping the distribution of matter on cosmic scales is nontrivial because $\sim 85\%$ of all the matter is dark: it does not interact with light and therefore we cannot observe it directly. One approach to do so is to use weak gravitational lensing (WL): as photons from far-away background sources (such as galaxies) travel toward us, the cosmic web itself acts as a gravitational lens, distorting their paths and thus distorting the shape of the sources.

Galaxies that are nearby will be lensed by similar structure so they will have correlated shapes. WL galaxy surveys look at millions of galaxies \cite{10.1093/mnras/stt2013} to find statistical correlations in their shapes. The large number of galaxies is needed both due to the intrinsic shape noise of galaxies as well as the weakly correlated nature of the underlying signal. These surveys construct shear maps, which can be used to reconstruct the projected distribution of matter in the universe between us and the sources. The resulting surface mass density maps are referred to as convergence maps. By providing a direct view into the distribution of dark matter across cosmic times, convergence maps can constrain cosmological parameters, the halo mass function, and can be cross-correlated with images at other frequencies to learn about halo bias and dependence of astrophysical processes on the dark matter density.

\subsection{Mock data}

In this work, we focus on the weak lensing convergence power spectrum. Like many other observables in cosmology that are the product of the highly nonlinear process of structure formation on small scales, writing down a tractable (and correct) likelihood function for weak lensing observables is very challenging. We can instead use simulations: given a set of cosmological parameters that we seek to constrain, we can run complex forward simulations to generate mock data, which we can then use to infer cosmological parameters from real data after extensive validation. Despite their computational cost, one advantage of using forward models is that many effects $-$ such as detector and theoretical systematics (e.g., intrinsic alignments and baryonic effects in the case of WL) $-$ are easier to incorporate into a forward model than into a likelihood.

In Appendix \ref{app:WL_maps} we detail how we obtain the mock convergence maps and power spectra used in this work. We summarize some key details here. 
We obtain 75,000 mock maps by running four $N$-body simulations with different initial seeds and using \texttt{LensTools} \cite{lenstools} to generate convergence maps from them. Each map has a sky coverage of 3.5 $\times$ 3.5 deg and 1024 $\times$ 1024 pixels. We set the density of matter $\Omega_{m}=0.3$, the density of dark energy $\Omega_{\Lambda} = 0.7$, the density of baryons $\Omega_{b} = 0.046$, the variance of matter fluctuations $\sigma_8 = 0.8$, the scalar spectral index $n_{\rm s} = 1$, and the Hubble constant $H_0 = 72$ km/s/Mpc. We obtain the convergence power spectrum in 34 different multipole bins uniformly distributed in log space for $\ell = [100,5000]$ (past $\ell = 5000$ the calculated power spectrum deviates from theory significantly \cite{cnn_wl1}). Figure \ref{fig:mock_data} displays an example of a mock convergence map (right) and the 68\% CL and 95\% CLs of the convergence power spectra in blue (left).

We point out that for the results presented in this paper we do not include observational effects in the maps such as noise or filtering. The reason for this is twofold. First, as we will show below, the NG is strongest at the lowest multipoles, where even a pessimistic amount of noise level would not Gaussianize the data. Furthermore,
while adding noise could Gaussianize high-$\ell$ bins, the objective of this work is to analyze the capacity of different DDLs to capture non-Gaussianities, and thus how strong the non-Gaussianities have to be in order to be adequately picked up, so we opt against introducing noise. In a follow-up paper where we study the impact of using FBLs on parameter inference, we take into account observational effects that would actually be present in the data.

\subsection{Weak lensing likelihood non-Gaussianity}

\begin{figure*} 
    \centering
    \begin{subfigure}[b]{0.8\textwidth}
        \includegraphics[width=\textwidth]{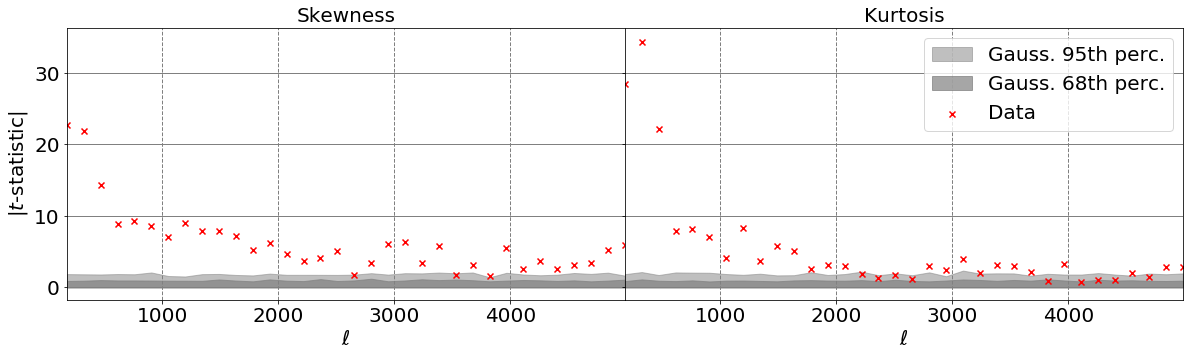}
    \end{subfigure}
    \begin{subfigure}[b]{0.8\textwidth}
        \includegraphics[width=\textwidth]{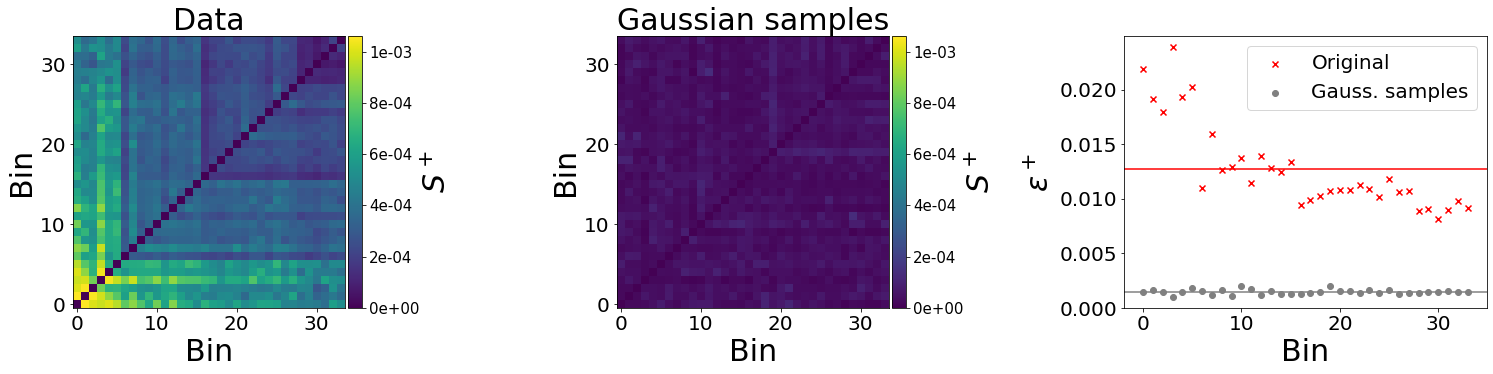}
    \end{subfigure}
    \begin{subfigure}[b]{0.3\textwidth}
        \includegraphics[width=\textwidth]{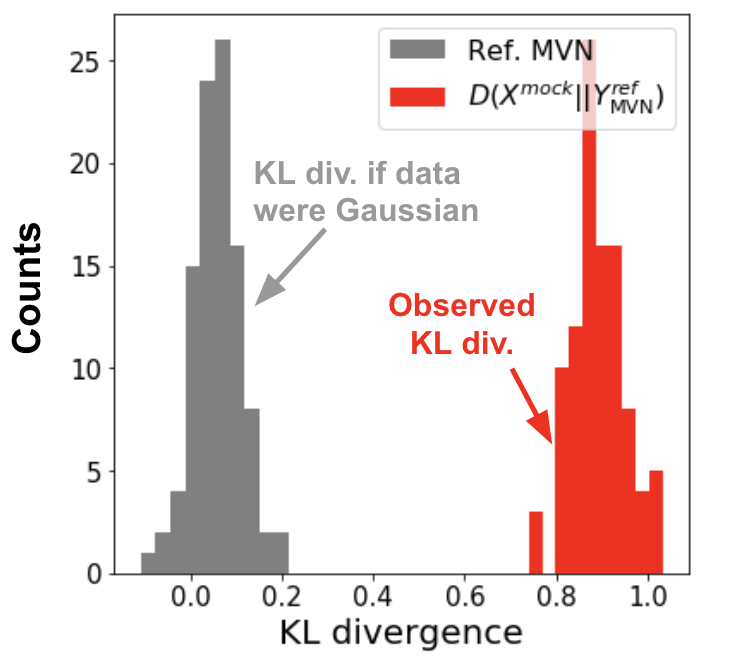}
    \end{subfigure}
    ~
\caption{\footnotesize{\textit{Top}: absolute value of the $t$-statistic of skewness (left) and kurtosis (right) of individual bins for 2,048 weak lensing mock power spectra in red. The gray contours correspond to averaging 100 different sets of 2,048 samples drawn from a multivariate normal likelihood with the same mean and covariance as the mock data. As expected, they correspond to $t$-statistic values of 1 (dark gray) and 2 (light gray). The fact that most of the red crosses lie above the Gaussian contours reflects the per-bin NG in the data. \textit{Middle}: the $S
^+$ matrix for the mock data (left) and equivalent Gaussian samples (middle). The sum over columns of each matrix, $\epsilon_u^+ \equiv \sum S^+_{u,v}$, is shown on the right as red crosses and gray circles, respectively. The red line is a factor of 8 higher than the gray line. Comparing the two $S^+$ matrices, there is structure in the data that is not present in the Gaussian samples. This, and the vertical offset between the red and gray crosses on the rightmost panel, reflect the pairwise NG in the data. \textit{Bottom}: nonparametric KL divergence estimate between the mock data and their Gaussian counterparts (red), and the Gaussian samples with themselves (gray). The fact that the gray histogram is not perfectly centered at zero is due to the slight variability of the KL estimator in $34$ dimensions, given the number of mocks considered (Appendix \ref{app:kldiv}). If the data were truly Gaussian, we would expect the red histogram to lie on (or near, see Appendix \ref{app:kldiv}) the gray histogram; the large horizontal offset reflects the NG in the distribution as a whole. }}\label{fig:WL_nongauss}
\end{figure*}

If a field is decomposed into spherical harmonics, with coefficients $a_{\ell,m}$, then an angular power spectrum bin at a given $\ell$ is given by:
\begin{equation}
    P_{\ell} = \frac{1}{2\ell +1} \sum_{m=-\ell}^{\ell} |a_{\ell,m}|^2.
\end{equation}
The transformation into $a_{\ell,m}$ space is linear, so Gaussian fields will have Gaussian-distributed $a_{\ell,m}$s. Because each bin is a sum over quadratic combinations of Gaussian-distributed variables, power spectrum bins are governed by Gamma distributions. The applicability of a Gaussian likelihood therefore decreases with decreasing multipole number (the sum is over less modes as $\ell$ decreases), which is precisely why CMB analyses only apply a Gaussian likelihood at high multipoles \cite{planck_2018_likelihood}. In WL analyses this distinction is not commonly made, and there are two additional effects that are important. The underlying field is itself non-Gaussian, meaning that the $a_{\ell,m}$s are not necessarily Gaussian-distributed. Furthermore, the fact that galaxies are discrete tracers (as opposed to a smooth random field like the CMB) also increases the skewness \cite{skewed_likelihood}. Finally, recall that, compared to the CMB, WL observables have the added complexity of requiring computationally-expensive forward-modeling.

Ref. \cite{Sellentin:2017fbg} showed that there are significant non-Gaussian correlations in the cosmic shear correlation functions (the real-space analogue of the power spectrum) for the weak lensing survey CFHTLenS \cite{cfhtlens}. Furthermore, they showed that because the one-point correlation statistics are skewed, WL datasets are likely to lead to a systematically low lensing amplitude. Since the WL amplitude increases with $\Omega_{m}$ and $\sigma_{8}$, the authors suggest that this could (partially) explain the discrepancy in the value of $S_8 = \sigma_8 \sqrt{\Omega_m}$ between WL and CMB surveys (the so-called $S_8$ tension \cite{S8_tension}). Finally, they show that the non-Gaussianities become more relevant on larger angular scales, meaning that this issue will be more relevant for upcoming wide-angle surveys such as \textit{Euclid} and LSST. This can be understood as a break of the CLT on these scales.

For our suite of NG tests (Section \ref{sec:nongauss}), we use the power spectra from an ensemble of 2,048 maps instead of the full set. The reason why we use this seemingly arbitrary number of mocks is to have a direct point of comparison to the caliber of non-Gaussianity discussed in the context of galaxy power spectra in Ref. \cite{Hahn:2018zja}: there, the authors only have access to 2,048 mocks for their observable. As we discussed above, our measures of non-Gaussianity are extensive (both in terms of number of bins and in terms of number of mocks), so an 
``apples to apples" comparison between two observables requires the same number of mocks and bins. We relegate the details of this comparison to Appendix \ref{app:BOSS}, but in short, we find that the WL power spectrum is significantly more non-Gaussian than the galaxy power spectrum at the scales considered, and its potential to have a larger effect on biasing inferred parameters makes it a more exciting target of data-driven likelihood methods. 

Finally, we pre-process the mock power spectra by subtracting off the mean and whitening them using the Cholesky decomposition of the precision matrix: $\Sigma^{-1} = \mathbf{L}\mathbf{L^{\rm T}}$, where $\Sigma$ is the covariance matrix and the superscript $T$ denotes the transpose. The data is whitened by applying the linear transformation $\mathbf{L}$ on the mean-subtracted mock data matrix. It can subsequently be unwhitened by applying the inverse transformation.

The results of the NG tests are shown in Figure \ref{fig:WL_nongauss}. The strong non-Gaussian signatures are apparent in all three tests. The top row shows the absolute value of the $t$-statistic for the distribution of each bin as red crosses. The dark (light) gray shaded region is the 1$\sigma$ (2$\sigma$) confidence level (CL) obtained from 2,048 Gaussian realizations drawn with mean and covariance extracted from the mock power spectra. As expected, the Gaussian CL matches a $t$ value of 1 (2). 

Clearly, the strongest NG correspond to the largest scales (lowest $\ell$), but notice that for nearly all bins the $t$-statistic is significantly larger than 2$\sigma$, and even 3$\sigma$. In Appendix \ref{app:tstats_panels_WL} we show the individual distributions for each bin, together with each KDE fit, which make the statistically significant skewness and kurtosis visible for many of them. 

The middle row shows the $S^+$ matrix for the mock data (left) and for Gaussian samples drawn with mean and covariance given by the mock data (center). The stark difference between both panels is easily visible by eye. The sum of each column in $S^+$, $\epsilon^+$, is shown on the right. The red crosses represent the mock data while the gray circles correspond to the Gaussian samples. There is a non-negligible gap between both, and in fact the mean of the red crosses (shown as a horizontal red bar) is $\sim 8$ times greater than the mean for the Gaussian mocks. 

Finally, the histograms in the bottom row reflect the nonparametric KL divergence test with the number of nearest neighbors set to $k=10$. Each histogrammed data point represents a KL divergence test between the 2,048 mock power spectra and 2,048 Gaussian power spectra drawn from an analogous MVN. We repeated this procedure 100 different times, with the red histogram showing the estimated KL divergence distribution. The gray histogram is the KL divergence of an ensemble of Gaussian mocks with another ensemble of Gaussian mocks, and therefore serves as a reference for the expected nonparametric KL divergence. The large horizontal offset between these two distribution reflects the fact that the true likelihood for the power spectra is not a multivariate Gaussian. 

\begin{figure*}
    \centering
    \includegraphics[width=\textwidth]{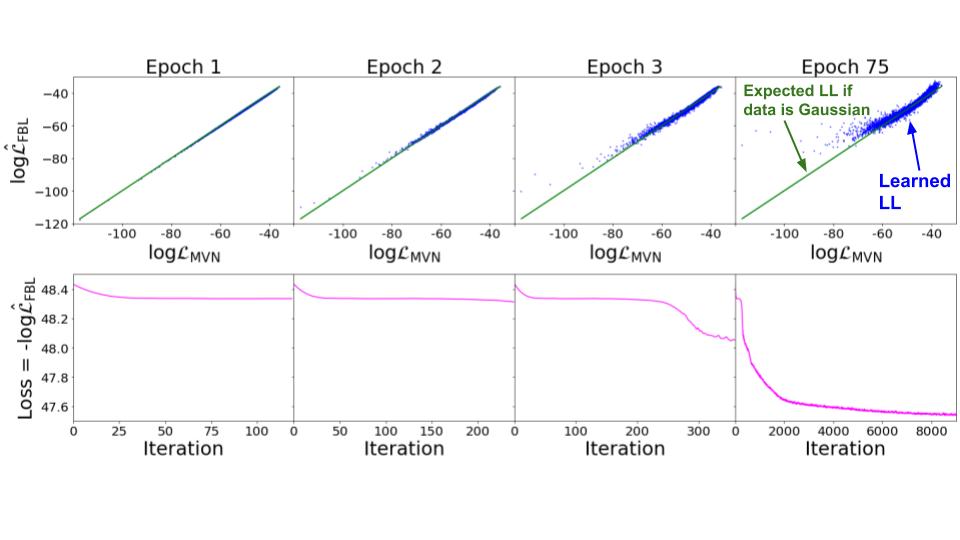}
    \caption{\footnotesize{\textit{Top}: In blue, log-likelihood of the test set samples under a multivariate normal likelihood versus the flow-based likelihood for various epochs. The green line is the MVN log-likelihood against itself, shown to make deviations from Gaussianity in the FBL more obvious.} \textit{Bottom}: test loss as a function of batch iteration number. Note that there are 120 iterations per epoch. } \label{fig:training}
\end{figure*}

\section{Learning Flow-Based Likelihoods (FBL)}\label{sec:apply_flow}

\subsection{Network architecture and training procedure}

Before training, we apply the same preprocessing steps as the ones mentioned in Section \ref{sec:wl_dset}: we subtract off the mean and whiten the data using the Cholesky decomposition of the precision matrix. Our tests on toy Gaussian data with a full rank covariance suggest that training benefits from standardizing the variance of each bin (see Appendix \ref{app:toy_problems}), both in terms of speed and in terms of the quality of the final likelihood fit. Furthermore, by destroying Gaussian correlations in the data, the network can focus on picking up non-Gaussian signatures, and by subtracting off the mean all bins are equally important to the network.

In terms of the network architecture, we stack a single continuous normalizing flow with a hidden layer of dimension $d=64$. We use exponential linear unit (ELU) activation functions, given by:
\begin{equation}
    \text{ELU}(\mathbf{x}) = \text{max}(0,\mathbf{x}) + \text{min}(0,\alpha *(\exp(\mathbf{x})-1)),
\end{equation}
with $\alpha = 1$ and $*$ denoting element-wise multiplication. The network has 13,449 parameters. 

Our mock dataset consists of 75,000 samples. We reserve 10\% for testing, 10\% for validation, and the remaining 80\% for training. We train the network with batches of 500 samples using the Adam optimizer \cite{adam} and a learning rate of 0.001. We do not introduce any regularization and thus rely on the validation loss to gauge overfitting. The results shown in the remainder of the paper correspond to training for 75 epochs.

We have checked that our results are robust to different activation functions, network architectures, and learning rate. Furthermore, we find that as the network is trained on $\gtrsim 20,000$ samples, it is able to learn the likelihood without overfitting (unsurprising, since as a rule of thumb the number of training samples has to be at least greater than the number of parameters in the model). 

Before training on real data, we analyzed the fidelity of FBLs on toy Gaussian data. We detail our results in Appendix \ref{app:toy_problems}, but summarize some of the relevant findings for training here. By studying the learned likelihood in Gaussian problems with singular and nonsingular covariance matrices, we noticed that although the sample quality was excellent in both, the recovered likelihood becomes significantly biased and imprecise as the determinant of the covariance approaches zero. Conversely, for a full-rank covariance the likelihood was recovered perfectly to within sampling error. Therefore, we tried whitening the data before training and found that indeed training is much faster and well behaved when we do so. 

\subsection{Results}

\begin{figure*}
    \centering
    \begin{subfigure}[b]{\textwidth}
        \includegraphics[width=0.8\textwidth]{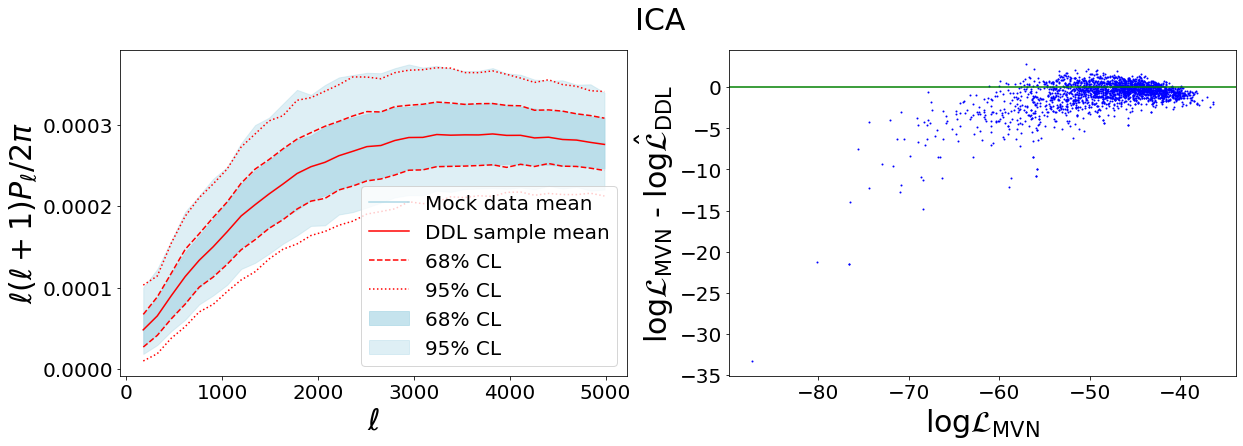}
    \end{subfigure}
    \begin{subfigure}[b]{\textwidth}
        \includegraphics[width=0.8\textwidth]{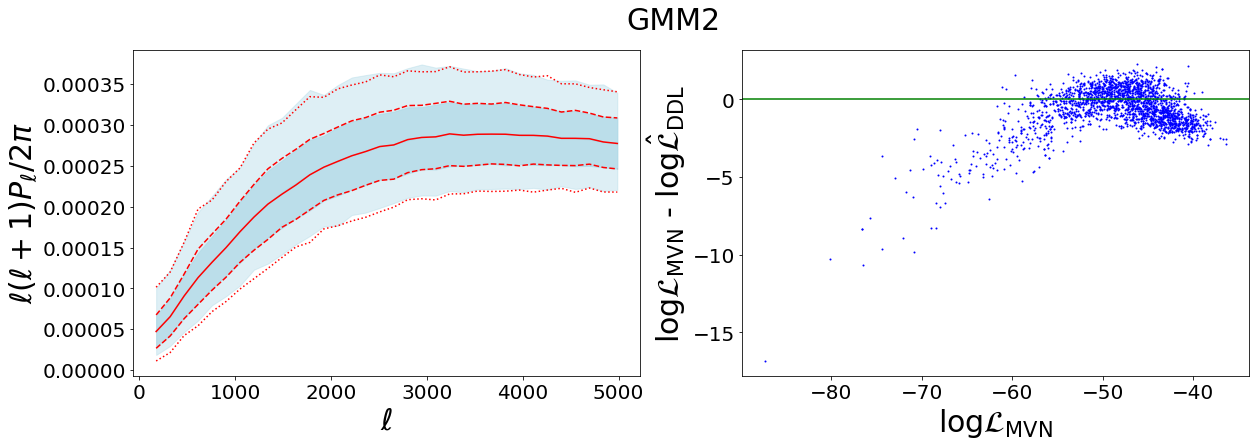}
    \end{subfigure}
    \begin{subfigure}[b]{\textwidth}
        \includegraphics[width=0.8\textwidth]{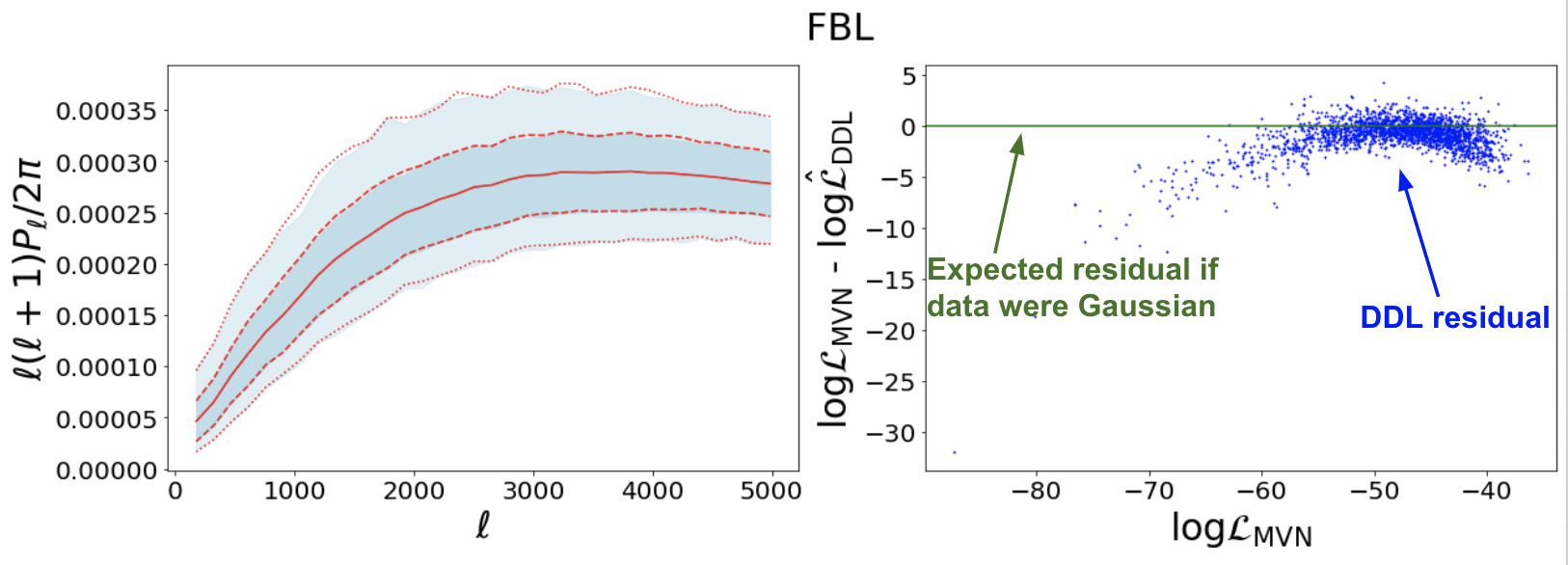}
    \end{subfigure}
    ~
\caption{\footnotesize{Each row corresponds to a different data-driven likelihood. From top to bottom: ICA, GMM2, and FBL. \textit{Left}: mock weak lensing convergence power spectrum 68\% and 95\% CLs (light blue) together with the CLs obtained from sampling the DDLs (red), as a function of multipole number $\ell$. \textit{Right}: residual between the log-likelihood of each test set sample under a MVN likelihood and under the DDL. While the samples drawn from all three DDLs appear to match the data when looking at the mean, 1 and 2$\sigma$ intervals, the likelihood is significantly different.}}\label{fig:DDL}
\end{figure*}

In Figure \ref{fig:training} we show the log-likelihood of the mock samples under a MVN likelihood versus that of the FBL (top), as well as the progression of the test loss as a function of iteration number (bottom). Note that we do not expect the FBL to match the likelihood under a MVN (quite the opposite, given the level of NG detected in Section \ref{sec:wl_dset}), we simply show it to see how the likelihood values are distorted with respect to the Gaussian likelihood commonly used for inference. Interestingly, we can actually see that after a full epoch FFJORD has learned the multivariate normal likelihood: the likelihood values of the test set under the FBL (blue) are in perfect agreement with the MVN likelihood (green line), and the loss plateaus due to having found a local minimum with the MVN likelihood. With further training, however, the network is able to pick up on the non-Gaussianities and we can see the loss starts decreasing while simultaneously the deviation away from the MVN likelihood becomes stronger in the test set. 

\begin{figure*}
    \centering
    \begin{subfigure}[b]{\textwidth}
        \includegraphics[width=0.9\textwidth]{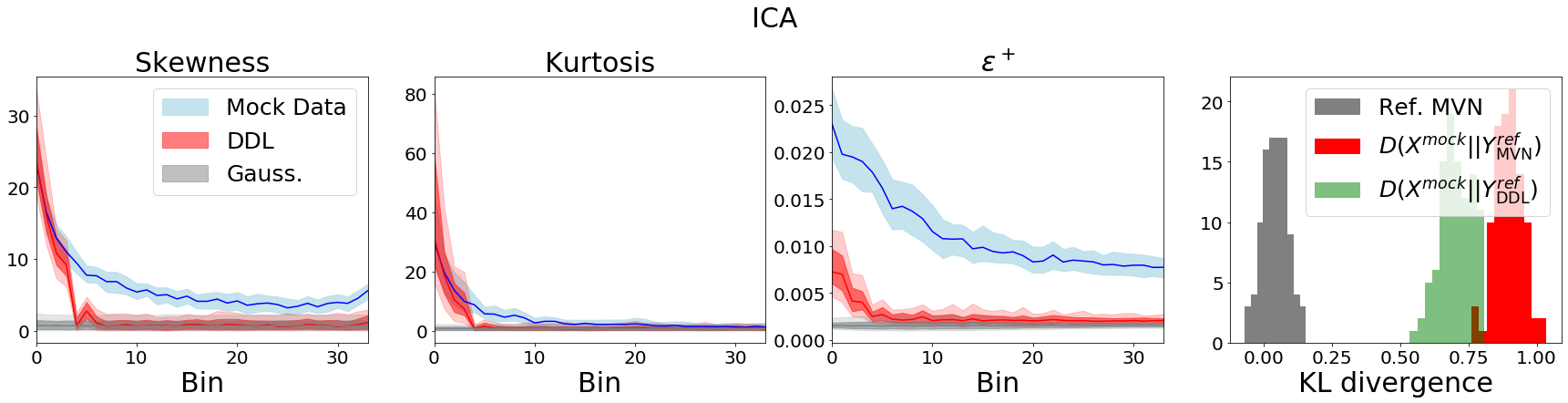}
    \end{subfigure}
    \begin{subfigure}[b]{\textwidth}
        \includegraphics[width=0.9\textwidth]{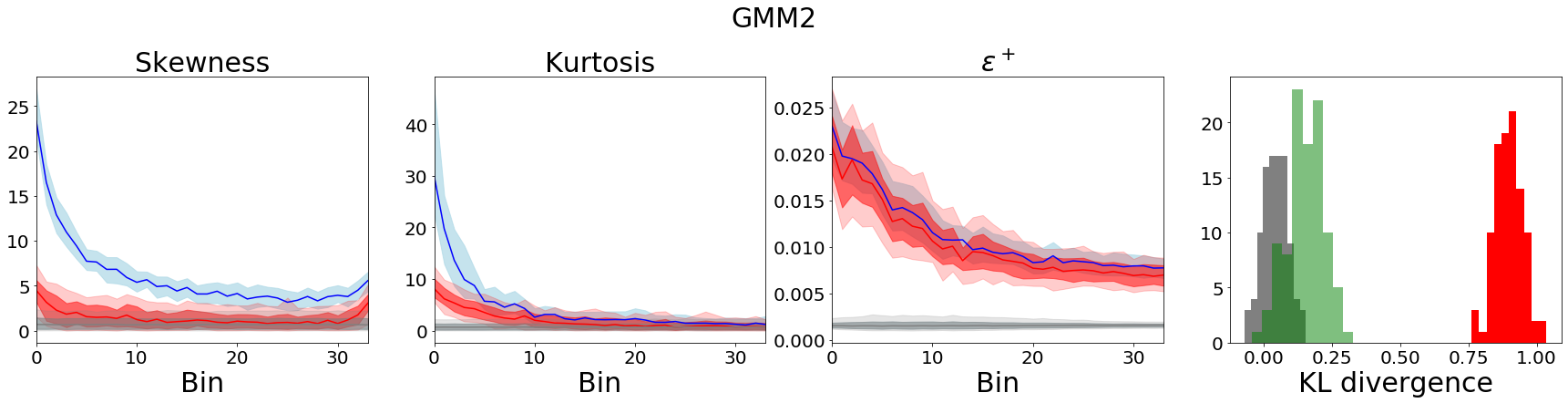}
    \end{subfigure}
    \begin{subfigure}[b]{\textwidth}
        \includegraphics[width=0.9\textwidth]{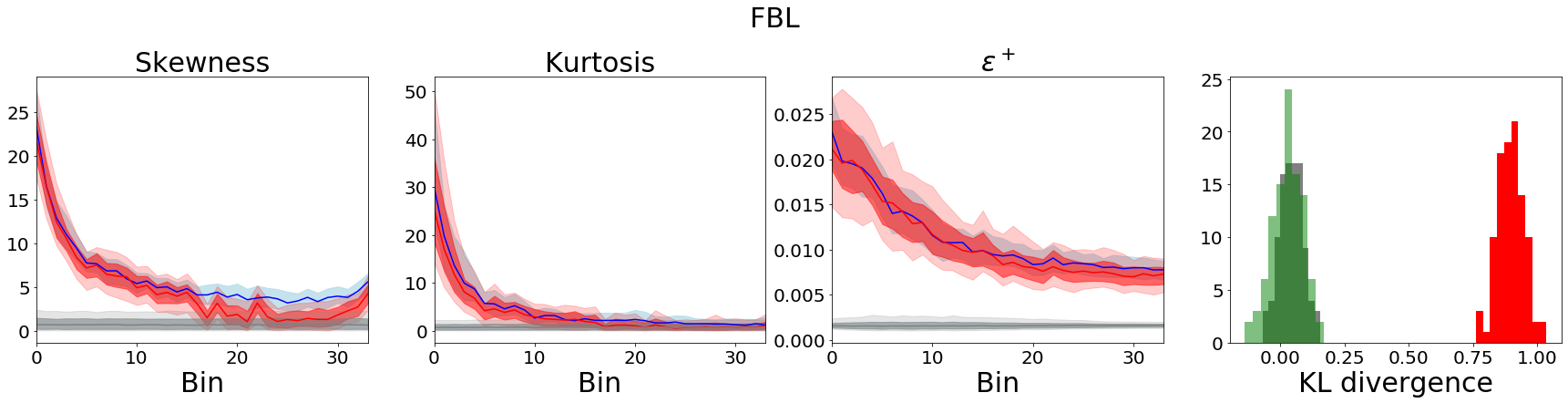}
    \end{subfigure}
    ~
\caption{\footnotesize{Each row corresponds to a different data-driven likelihood. From top to bottom: ICA, GMM2, and FBL. From left to right: absolute value of the $t$-statistic of the skewness, kurtosis, and $\epsilon^+$ of each bin for 100 different sets of 2,048 mock WL power spectra (blue), DDL samples (red), and Gaussian samples (gray). The vertical offset between the blue and gray shaded regions reflects the high NG in the data. Overlap between the red and blue contours indicates that the DDL has captured the NG. The rightmost panel shows the KL divergence between a Gaussian with itself (gray), between the data and the MVN likelihood (red), and between the data and the DDL likelihood (green). The horizontal offset between the red and gray histograms reflects the fact that the true likelihood is not MVN. If the DDL has captured the true likelihood correctly, we expect the green histogram to lie on (or near, see Appendix \ref{app:kldiv}) the gray histogram. }}\label{fig:DDL_samples}
\end{figure*}

We compare the FBL to two other DDLs: ICA with 34 components (i.e. no dimensionality reduction) and a Gaussian mixture model with two components (chosen by minimizing the Bayesian information criterion; from here onwards, we refer to this model as GMM2). Note that we fit these two likelihoods with the full set of weak lensing mocks as well, to ensure that the comparison to the FBL was fair. In Appendix \ref{app:data_lim} we show how our results are affected by fitting the DDLs with many less mocks, to mimic the data-limited regime in which some previous works that have applied DDLs in cosmology have found themselves in. In short, we find that the DDLs severely underestimate the non-Gaussianities in the data, concluding that the claimed parameter shifts works that have applied DDLs in this regime have observed are unlikely to truly incorporate the full extent of the impact that non-Gaussinities can have, if modeled correctly. 

In Figure \ref{fig:DDL}, each row corresponds to a different DDL likelihood: ICA, GMM2, and FBL from top to bottom. The left column shows the true mean, 68\% and 95\% CLs of the mock data in blue, and the same quantities obtained from sampling the DDL and unwhitening in red. The $1\sigma$ contours of all likelihoods show excellent agreement with the data, while the 2$\sigma$ boundaries show small discrepancies, especially for ICA.

The right-hand panels show the residual between the MVN likelihood and the DDL for the test set samples. The nonzero residual for most test set samples indicates that the DDLs are picking up on some NG features in the data. All three DDLs show a similar moon-shaped residual, although it is more apparent for GMM2 and FBL than for ICA. This shows that the MVN likelihood is not capturing the tail-ends of the data distribution correctly, which could have important implications in inference pipelines. 

By generating samples from all three likelihoods we can also carry out the three non-Gaussianity tests detailed in Section \ref{sec:nongauss}. In Figure \ref{fig:DDL_samples}, the red contours in the first three columns correspond to the 68\% and 95\% CLs of the skewness, kurtosis, and $\epsilon^+$, from left to right, from drawing sets of 2,048 samples from each likelihood 100 different times. They can be contrasted to the ones obtained from 100 different draws of 2,048 power spectra from the full set of mocks, which are shown in blue, and to the ones that correspond to draws from a MVN, shown in gray. Furthermore, the fourth column shows the KL divergence between the mock data and a MVN likelihood (red); between the mock data and the DDL (green); and between a MVN and itself (gray). 

ICA and GMM2 each seem to have different strengths. ICA is able to pick up on the strong skewness and kurtosis of the first 2-4 bins, since they are the ones that display the strongest non-Gaussianities. On the other hand, the GMM2 approximates the true $\epsilon^+$ while failing to capture most of the skewness and kurtosis. Interestingly, by looking at the right-hand panels of these two DDLs, we can see that capturing the pairwise NGs is much more strongly correlated to improving the KL divergence of the distribution as a whole. The shortcomings of GMM2 in terms of the skewness and kurtosis are overcome by the FBL, which shows excellent agreement with the data except the skewness at the highest multipoles, and the KL divergence estimate between the FBL and the data (green) has near-perfect alignment with the reference distribution (gray). 

\section{Discussion}
\label{sec:discussion}

We have shown that, for our mock weak lensing data, neither the GMM nor the ICA likelihoods fully succeed in capturing the NG in the data, while the FBL does so extremely well. By considering the underlying principles behind each of the three DDLs we can try to understand their performance when applied to the task of capturing non-Gaussianities. 

The fact that ICA is predicated upon maximizing the non-Gaussianity of the components means that it is able to reproduce the distributions of independent bins, even when these have significant skewness or kurtosis. However, because it also requires independence between dimensions, it destroys the non-Gaussian correlations between datapoints that the $S^+$ test is sensitive to. On the other hand, GMMs cannot account for strong distortions away from Gaussianity of individual bins. The FFJORD-based FBLs are built from transformations with unconstrained Jacobians, and are thus given much more freedom of expressivity than the other DDLs considered in this work. Nevertheless, although the FBL outperforms the ICA and GMM likelihoods in terms of the three NG tests carried out, we find that there is a minimum threshold for the per-bin NG below which the FBL struggles to distinguish from a Gaussian (as can be seen for the skewness at high mutipoles). 

The strengths and weaknesses of each of the three DDLs offers strong evidence that data volume is not the only factor that will determine the success or failure of a DDL. Ultimately, having some understanding of the type of NG present in the data is crucial to select the right model and not underestimate the impact of NG when inferring parameters from the data. Using our multi-resolution NG tests, which focus on increasingly coarser levels of non-Gaussianity in the data, is beneficial to faithfully diagnose the quality of the DDLs: in isolation, they could mislead one into having false confidence in the learned likelihood (like the bin-wise non-Gaussianity for ICA or $\epsilon^+$ for GMM2), but taken together they succeed in identifying shortcomings in each of these models. The fact that the FBLs succeed in capturing the different types of NG diagnosed through the three tests suggests that FBLs are likely to be widely applicable across datasets and domains much more readily than ICA and GMM, which can require a trial-and-error procedure \cite{Hahn:2018zja}.

The impact of using FBLs for parameter inference, instead of MVN or another DDL, is left for a follow-up paper, but looking at the deviations from the Gaussian expectation can give us some insight. By comparing the evolution of the loss during training to the likelihood of the samples, we can see how the loss is progressively minimized as the likelihood of the test samples deviates from the Gaussian expectation. The trained model shows that the FBL boosts the lowest likelihood values the most, although it also boosts the highest ones. This suggests that, at the very least, misuse of MVN likelihoods could be underestimating uncertainties in inferred parameters.

\section{Conclusion}
\label{sec:conclusion}

In this paper, we have investigated the use of data-driven likelihoods to capture non-Gaussianities in the data. In particular, we have suggested exploiting flow-based machine learning models. These models are interesting because the loss function used for optimization is the negative log-likelihood of the data itself. We focus on the quality of this optimized likelihood and its capacity to pick up non-Gaussian signatures in the training data, with the goal of using it for inference. We refer to it as flow-based likelihood, or FBL. 

We applied the FBL to a  significantly non-Gaussian mock cosmological dataset: the weak lensing convergence power spectrum. We built on the work of Refs. \cite{Hahn:2018zja,Sellentin:2017fbg} to design a suite of tests that seek to capture different non-Gaussian features in the data $-$ from a bin at a time to the observable as a whole $-$ and used it to gauge to what extent three different data-driven likelihood methods succeed in capturing the non-Gaussianities in the data they are fit on. We used two DDLs used in Ref. \cite{Hahn:2018zja} for different cosmological large-scale structure observables, ICA and GMM, and contrasted them to our proposed FBL. An interesting point to keep in mind is that the non-Gaussianities exhibited by the mock weak lensing data are much stronger than the ones of the mock galaxy power spectra used in that work, meaning that while they found that using DDLs lead to small posterior shifts compared to a Gaussian likelihood, the shift could be greater for a dataset such as this one (see also Appendices \ref{app:data_lim}, \ref{app:kldiv}, and \ref{app:BOSS}).

We found that the FBL captured the underlying likelihood much better than the other two DDLs we considered: neither GMM nor ICA fully succeeded in capturing a vast portion of the non-Gaussianities. Through our three NG tests we were able to gain some insight into the applicability of each of these three DDLs, showing that the NG structure in the data can determine whether a given DDL is appropriate or not. Seeing the strong impact of pairwise non-Gaussian correlations in WL data, and the shortcomings of ICA in addressing them, is particularly interesting since works such as Ref. \cite{cnn_wl1} used an ICA dimensionality reduction before performing inference on weak lensing data, and concluded that the impact of a non-Gaussian likelihood was small. 

Unlike GMM and ICA, which require some restrictions in order to fit the data (e.g., independence between components in ICA), the FBLs used in this work are the product of transformations with unconstrained Jacobians, which allows them to be much more expressive. Not only do they succeed in capturing the NG in the data, but this freedom makes it likely for them to be widely applicable across datasets with different types of NG. Furthermore, the flexibility of the FBL when fitting different types of NG could prove beneficial not only to prevent having to follow a trial-and-error procedure to find appropriate DDLs for different observables, but also to avoid choosing a wrong one altogether.

One final consideration when weighing what DDL to apply is also the quantity of (mock) data available for training. While all data-driven methods are data-intensive, out of the three methods we consider, ICA requires significantly less parameters than GMMs, and GMMs than FBLs. Thus, depending on the computational expense required to generate mock data, employing a FBL could be prohibitive. In such a setting, looking at the type of NG features in the data through an approach like the one we suggest could aid the selection of an adequate DDL that is more restrictive than an FBL but requires less data.

When placing this work in the context of other research that has studied the impact of non-Gaussianity in WL data in particular, we want to point out that data is sometimes ``Gaussianized", by for instance combining bins) 
%or projecting the data into the few principal components with the highest signal-to-noise) 
before applying Gaussian and/or non-Gaussian likelihoods. Such works have concluded that NGs do not shift posteriors considerably (e.g., \cite{lin2019nongaussianity,delfi2}). 
However, this process can destroy potentially-useful information, and we therefore advocate not Gaussianizing the data and instead opting for more accurate non-Gaussian likelihoods, such as FBLs. This plight is addressed in the context of the thermal Sunyaev-Zel'dovich one-point PDF in Ref. \cite{act}, which discusses the unfortunate need to Gaussianize clearly non-Gaussian data due to not having access to an adequate non-Gaussian likelihood, with the consequence of weakening the constraining power of the PDF. We emphasize that even highly complex and data-driven approaches can (inadvertently) Gaussianize the data in their pipelines (e.g., \cite{Alsing_2019}), and thus advocate for scrutinizing data processing steps to ensure that non-Gaussian information isn't erased.

This work has required the use of simulated data to fit the three DDLs. The use of mocks in canonical large scale structure analyses is widespread, usually to estimate covariance matrices. As we have previously emphasized, adding theoretical and observational systematics can be easier with mocks than it is to incorporate directly into a likelihood. Mocks used for inference are extensively validated against the data they are needed for, but there is always the possibility that the way in which these effects are incorporated into mocks is insufficient or incorrect. Just like this could lead to an incorrect covariance being extracted from mocks, it can lead to DDLs that miss certain effects that are pertinent to the data. Understanding whether this deficit can be more important for a machine learning-based DDL such as FBLs, given that domain adaptation is nontrivial, than it is for other DDLs is left to future work.

We emphasize that this method is certainly not restricted to cosmological data. Inference in any domain can be improved by relaxing assumptions about the Gaussianity of the data. In particular, there is nothing about FBLs or the NG pipeline in this work that is specific to cosmology, and can thus easily be applied in any domain. Nevertheless, there are intriguing tensions in cosmological data that have caused much interest in the community \cite{S8_tension,Tensions,features} and, thus, make it an exciting target of FBLs. By relying on the Gaussian approximation for inference, we might be biasing inferred parameters or being falsely confident in them. While in the past this approximation may have been sufficient, as data precision increases considerably we enter a regime in which errors induced by using incorrect likelihoods can be significant.

\acknowledgements

C.D. was partially supported by Department of Energy (DOE) grant No. DE-SC0020223. We thank Elena Sellentin and ChangHoon Hahn for useful discussions, and Bryan Ostdiek, Jasper Snoek and Josh Speagle for feedback on a previous draft. In addition to the software cited throughout the text, this work would not have been possible without a wide array of publicly available \texttt{Python} packages, most notably: \texttt{NumPy} \cite{numpy}, \texttt{SciPy} \cite{scipy}, \texttt{PyTorch} \cite{pytorch} and \texttt{Matplotlib} \cite{matplotlib}.

\appendix 

\section{Likelihood versus sampling density} \label{app:likelihood_vs_sampling_density}

Throughout this paper, we have referred to the sampling density learned from the data, $\hat{p}_{\rm DDL}(\mathbf{x})$, as a learned likelihood, although these are distinct entities. In this Appendix we explain the reason behind our use of this terminology, and reconcile the difference between these two functions. 

For some data $\mathbf{x}$, the likelihood is a function of the model parameters $\boldsymbol{\theta}$: $\mathcal{L}_{\mathbf{x}}(\boldsymbol{\theta})$. The likelihood is not a probability density function, and as such does not have the same restrictions (e.g., having to integrate to unity in the case of a continuous distribution). Conversely, for a choice of parameters $\boldsymbol{\theta}$ the probability density function is in fact normalized to unity when integrated over the data: $\int d\mathbf{x} p_{\boldsymbol{\theta}}(\mathbf{x}) = 1$. Although they have the exact same functional form (up to constant factors), they represent different things: the likelihood is a function of the model parameters with the data fixed, while the PDF is a function of the data with the model parameters fixed. 

In this paper, for the DDL methods discussed we used $N$ mocks for a given value of the cosmological model parameters $\boldsymbol{\theta}$. This is analogous to likelihood analyses that use the mocks for a given cosmology to obtain the covariance matrix and then employ a MVN likelihood:

\begin{equation}\label{eq:mvn_like}
    p(\mathbf{x}|\boldsymbol{\theta}) \propto \exp\left[(\mathbf{x} - \boldsymbol{\mu}(\boldsymbol{\theta}))^T \widehat{\Sigma}^{-1} (\mathbf{x} - \boldsymbol{\mu}(\boldsymbol{\theta})) \right],
\end{equation}

\noindent where we have put a hat on the precision matrix to emphasize that it is an estimated quantity. This means that $\Sigma$ is evaluated at some fiducial cosmology and assumed to be cosmology-independent. In principle, $\Sigma$ does vary with cosmology, and this dependence can have a significant impact on inferred parameters \cite{refId0,Morrison_2013,White_2015}. 

Note that the dependence on the cosmological parameters enters the likelihood through the estimate of the mean $\boldsymbol{\mu}(\boldsymbol{\theta})$. We can rewrite Eq. (\ref{eq:mvn_like}) as 
\begin{equation}\label{eq:mvn_like_delta}
    p(\mathbf{x}|\boldsymbol{\theta}) \propto \exp\left[\Delta(\boldsymbol{\theta})^T \widehat{\Sigma}^{-1} \Delta(\boldsymbol{\theta}) \right],
\end{equation}
to emphasize that the parameter dependence is encoded in the difference between the data and the model $\Delta(\boldsymbol{\theta})$. To find the parameters that provide the best fit for the data, it is necessary to have a model to obtain $\boldsymbol{\mu}$ for any point in the allowed parameter space.

In this paper, our estimates from mock data are analogous to the step of estimating the covariance matrix in the MVN example above in the sense that all the mocks correspond to a single value of the cosmological parameters. Unlike in the MVN case, what we estimate from the mocks is the full sampling density $\hat{p}_{\rm DDL}(\mathbf{x})$. However, we interchangeably use the term DDL to refer to this function because it can in fact be used as a likelihood \cite{Hahn:2018zja}. All we require is an array $\Delta(\boldsymbol{\theta})$ and we can apply $\hat{p}_{\rm DDL}$ as a likelihood, which will now depend on the cosmological parameters: $\hat{p}_{\rm DDL}(\boldsymbol{x}|\boldsymbol{\theta})$. Thus, the terminology data-driven likelihoods in this paper foreshadows the possibility of using them as true likelihoods for inference. Although the focus here was showing their ability to capture NG in the data, using them as actual likelihoods is the ultimate goal of developing this method, and is done in a forthcoming paper. 

We emphasize that, just like in the case for a MVN likelihood, where using a fixed-cosmology covariance matrix is an approximation made for convenience (otherwise every point in the parameter space of $\boldsymbol{\theta}$ requires a whole suite of mock data), the impact of using a DDL estimated with mock data from a single cosmology has to be studied. The cosmology dependance will also be addressed in a forthcoming paper.

\section{Toy problems}\label{app:toy_problems}

\renewcommand{\thefigure}{B\arabic{figure}}

\setcounter{figure}{0}

\begin{figure*}
    \centering
    \begin{subfigure}[b]{0.75\textwidth}
        \includegraphics[width=\textwidth]{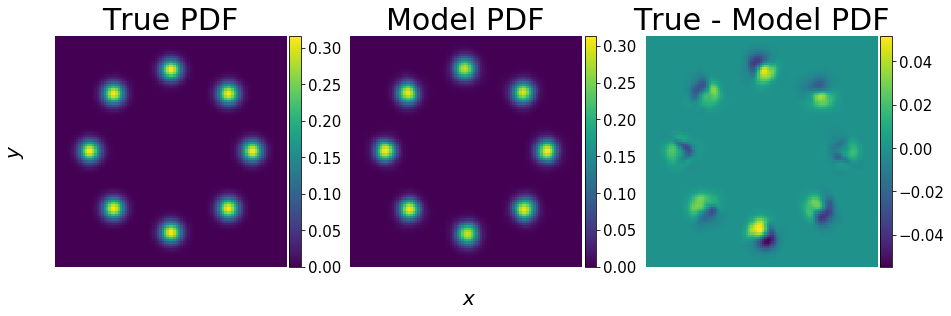}
    \end{subfigure}
    \begin{subfigure}[b]{0.75\textwidth}
        \includegraphics[width=\textwidth]{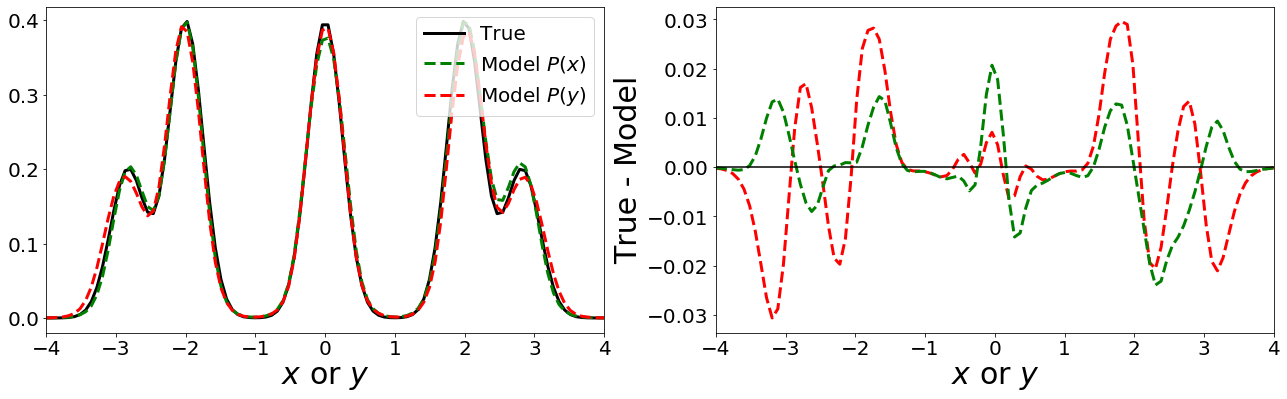}
    \end{subfigure}
    \begin{subfigure}[b]{0.3\textwidth}
        \includegraphics[width=\textwidth]{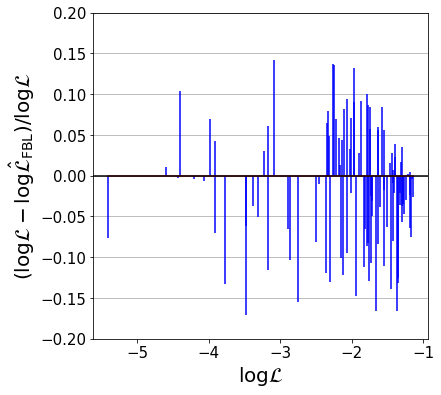}
    \end{subfigure}
    ~
\caption{\footnotesize{ \textit{Top}: true (left) and flow-reconstructed PDF (middle), together with their residual (right). \textit{Middle}: true and reconstructed marginal PDF (left) along $x$ (red) or $y$ (green), together with their residual (right). \textit{Bottom}: fractional difference between the true log-likelihood $\mathcal{L}$ and the FBL $\hat{\mathcal{L}}$ as a function of the true log-likelihood.}}\label{fig:2D_data}
\end{figure*}

To study the quality of the likelihoods provided by a flow-based model such as FFJORD, we first analyze the results with data whose likelihood is known and we can sample from. The advantage of these toy problems is that (1) we can compare the likelihood given by the model and the true likelihood and (2) by being able to sample from the likelihood, we can obtain a virtually infinite number of unique training samples. 

\subsection{In two dimensions}

Our first stepping stone is simple two-dimensional data. The benefit in two dimensions is that we can visualize the reconstructed PDF. We train FFJORD on a dataset whose PDF is comprised of 8 equal, symmetric Gaussians arranged around a circle. We use three stacked continuous normalizing flows with hidden dimensions $d=256$.

The results are shown in Figure \ref{fig:2D_data}. The top left panel is the true PDF, while the middle panel is the one obtained after training FFJORD. In Ref. \cite{ffjord}, the authors also showed these two panels. What interests us is the difference between them and how this translates into log-likelihood values. The top right panel shows the residual, which is on the order of 10\%. The middle row shows the same but for the marginalized probability distributions along the $x$ and $y$ axes. Finally, the bottom row shows the fractional log-likelihood residual between the true log-likelihood $\mathcal{L}$ and the model log-likelihood $\hat{\mathcal{L}}$. We can see that the scatter is around 10-15\%.

\begin{figure*}
\captionsetup[subfigure]{justification=centering}
    \centering
    \begin{subfigure}[b]{0.72\textwidth}
        \includegraphics[width=\textwidth]{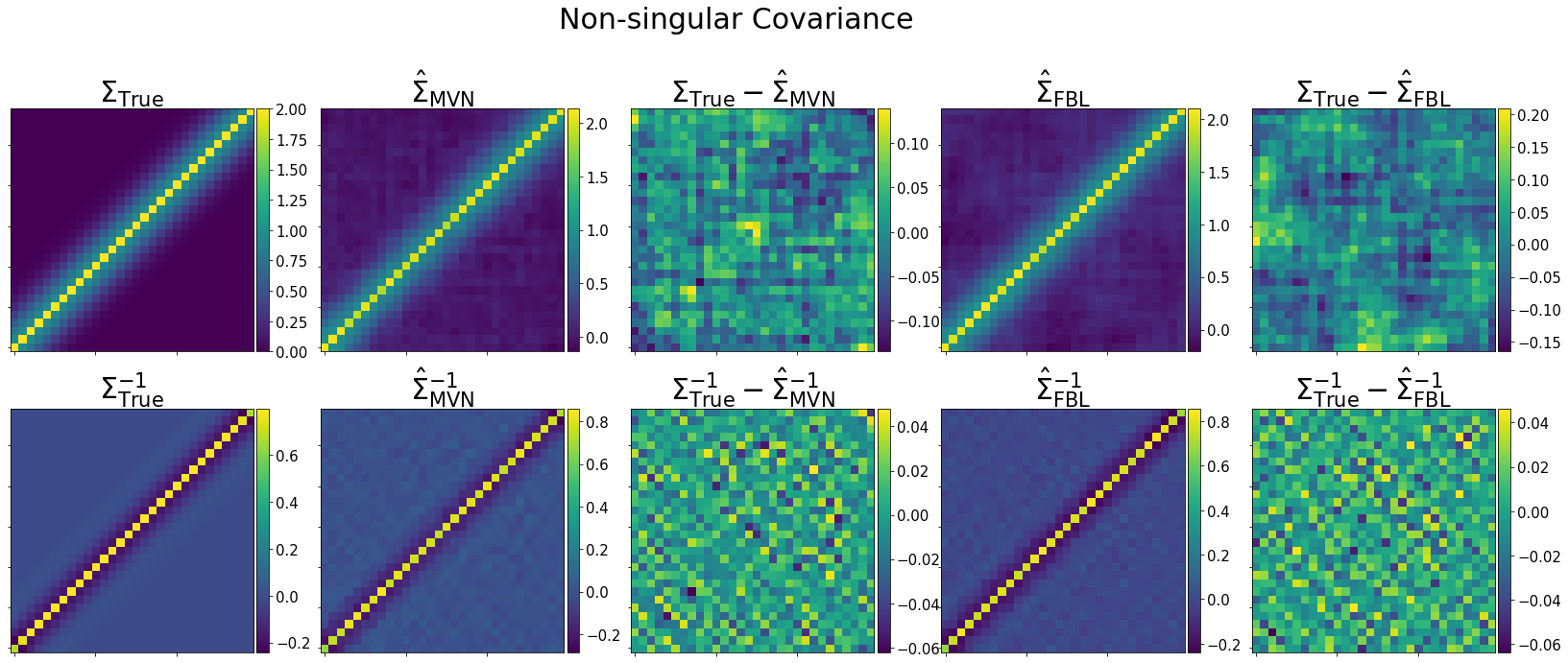}
        \subcaption{}
    \end{subfigure}
    \begin{subfigure}[b]{0.72\textwidth}
        \includegraphics[width=\textwidth]{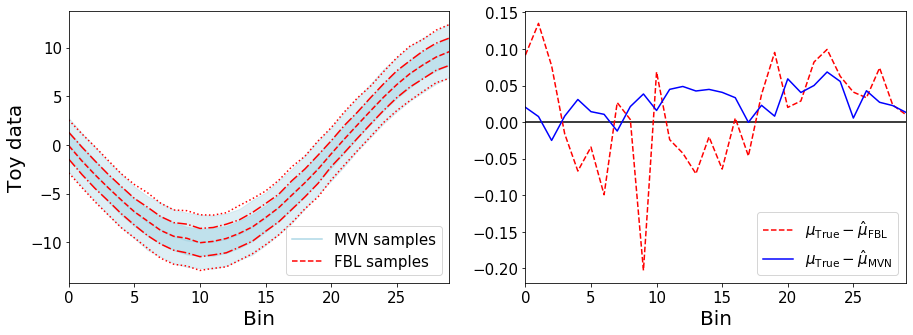}
        \subcaption{}
    \end{subfigure}
    \begin{subfigure}[b]{0.72\textwidth}
        \includegraphics[width=\textwidth]{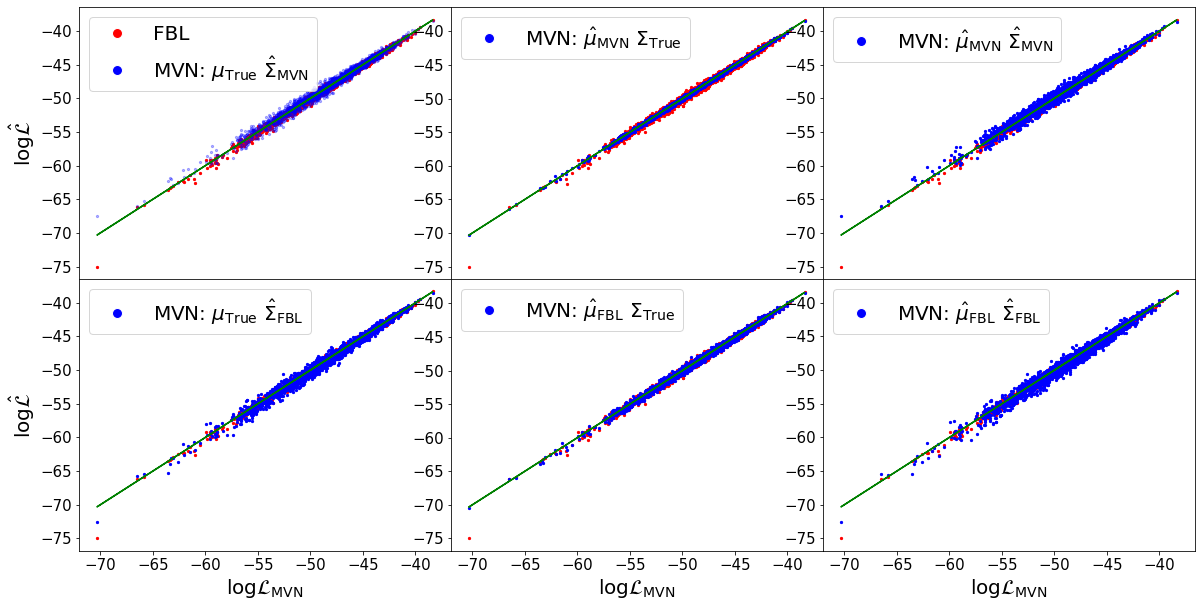}
        \subcaption{}
    \end{subfigure}
    ~
\caption{\footnotesize{Results when training FFJORD on Gaussian 30-dimensional data with mean and covariance given by Eqs. (\ref{eq:mu}) and (\ref{eq:K}), and noise added to the covariance to make it full rank. (a): The true covariance $\Sigma_{\rm True}$ and precision matrix $\Sigma_{\rm True}^{-1}$, together with the MVN sample covariance $\hat{\Sigma}_{\rm MVN}$ and precision $\hat{\Sigma}_{\rm MVN}
^{-1}$,  and FBL sample covariance $\hat{\Sigma}_{\rm FBL}$ and precision $\hat{\Sigma}_{\rm FBL}^{-1}$, obtained from 2,000 samples. It can be seen that the FFJORD-reconstructed matrices are accurate to within sampling error. (b): On the left, the mean, 68\% and 95\% CLs for the toy data in blue, and for 2,000 FBL samples in red. On the right, the difference between the true mean and the FBL sample mean (red) and between the true mean and MVN sample mean (blue). The FBL mean, 1 and 2$\sigma$ contours match the data very well; the mean residual is of the same order of magnitude as the error due to sampling. (c): the green line corresponds to the true log-likelihood of the test data. In blue, the log-likelihoood of the test data under a MVN for different combinations of the mean and covariance in blue. In red, the log-likelihood of the test data under the FBL. Agreement between the red scatter points and the green line would reflect that the FBL has learned the likelihood perfectly. The small scatter around the green line can be attributed to sampling error, as it is also present when using a sample mean and covariance with a MVN likelihood. }}\label{fig:toy}
\end{figure*}

\begin{figure*}
\captionsetup[subfigure]{justification=centering}
    \centering
    \begin{subfigure}[b]{0.85\textwidth}
        \includegraphics[width=\textwidth]{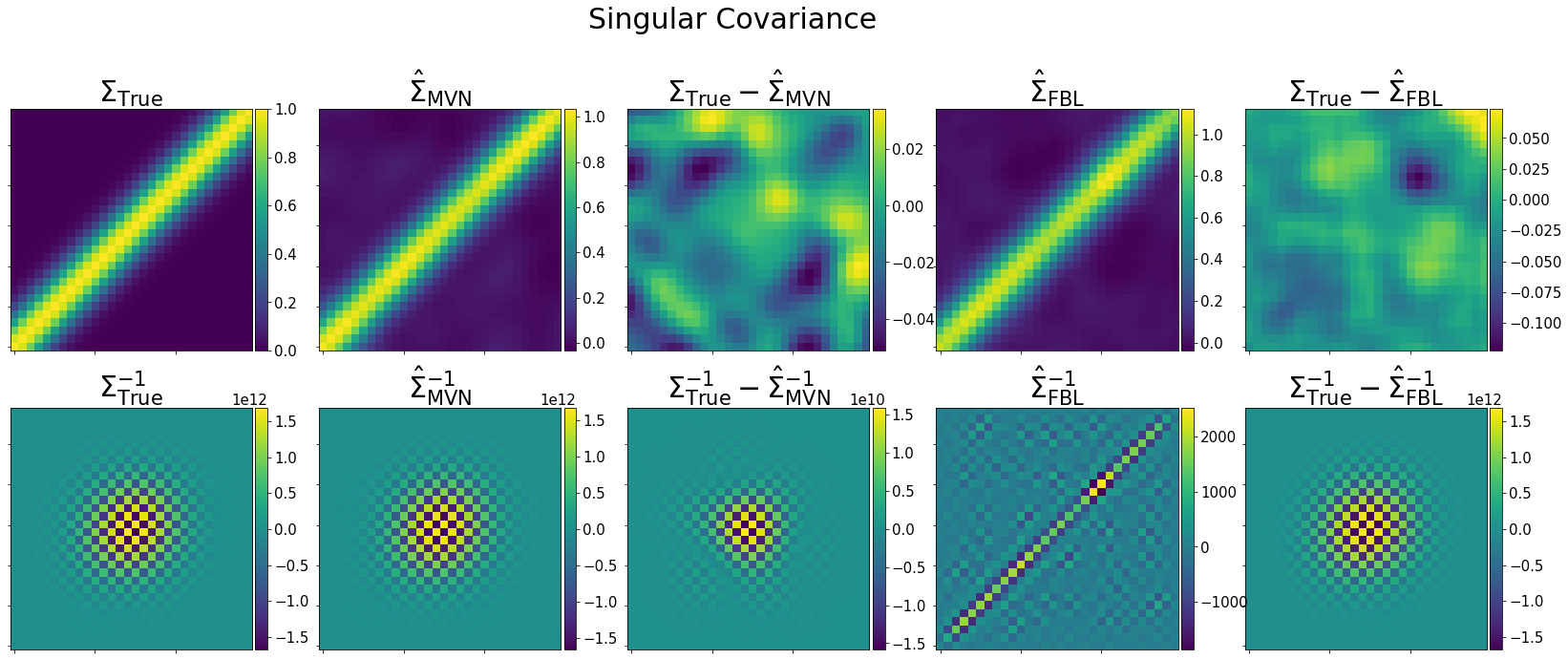}
        \subcaption{}
    \end{subfigure}
    \begin{subfigure}[b]{0.75\textwidth}
        \includegraphics[width=\textwidth]{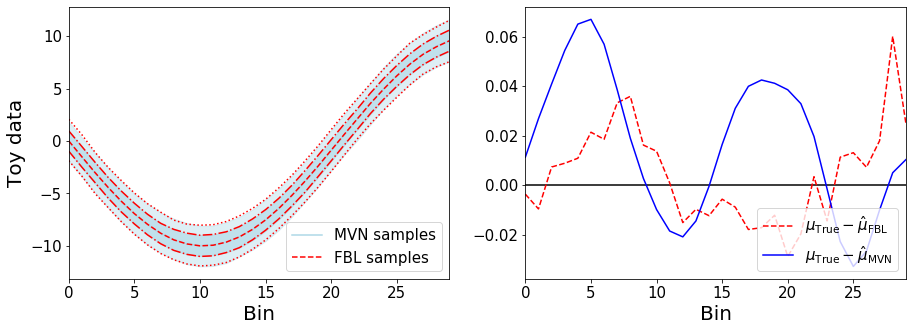}
        \subcaption{}
    \end{subfigure}
    \begin{subfigure}[b]{0.75\textwidth}
        \includegraphics[width=\textwidth]{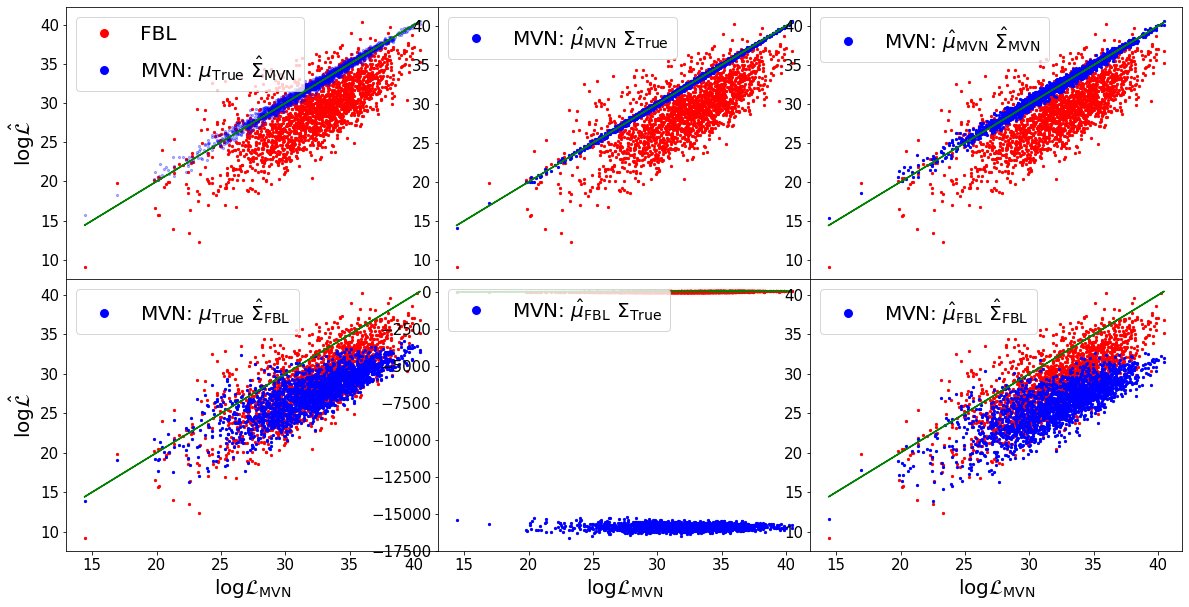}
        \subcaption{}
    \end{subfigure}
    ~
\caption{\footnotesize{Results when training FFJORD on Gaussian 30-dimensional data with mean and covariance given by Eqs. \eqref{eq:mu} and \eqref{eq:K}. (a): The true covariance $\Sigma_{\rm True}$ and precision matrix $\Sigma_{\rm True}^{-1}$, together with the MVN sample covariance $\hat{\Sigma}_{\rm MVN}$ and precision $\hat{\Sigma}_{\rm MVN}
^{-1}$,  and FBL sample covariance $\hat{\Sigma}_{\rm FBL}$ and precision $\hat{\Sigma}_{\rm FBL}^{-1}$, obtained from 2,000 samples. (b): On the left, the mean, 68\% and 95\% CLs for the toy data in blue, and for 2,000 FBL samples in red. On the right, the difference between the true mean and the FBL sample mean (red) and between the true mean and MVN sample mean (blue). (c): the green line corresponds to the true log-likelihood of the test data. In blue, the log-likelihoood of the test data under a MVN for different combinations of the mean and covariance in blue. In red, the log-likelihood of the test data under the FBL.}}\label{fig:singular_toy}
\end{figure*}

\subsection{In higher dimensions}

The next step is gauging the flow-based likelihood obtained in higher dimensions, on the order of the mock observables discussed in the main text. We train FFJORD on samples drawn from a 30-dimensional Gaussian distribution with mean 
\begin{equation}\label{eq:mu}
    \mu = 10 \sin(x)
\end{equation}
and a covariance matrix obtained from a squared exponential kernel $K$:

\begin{equation}\label{eq:K}
K(x,x') = \sigma^2 \exp\left( -\frac{(x - x')^2}{2 L^2} \right),
\end{equation}

\noindent where $\sigma$ is the standard deviation and $L$ the correlation lengthscale. We set $\sigma=1$ and $L = \sqrt{8}$. Note that this choice of mean and covariance was completely arbitrary.

\renewcommand{\thefigure}{C\arabic{figure}}

\setcounter{figure}{0}

\begin{figure*}
    \centering
    \includegraphics[width=0.9\textwidth]{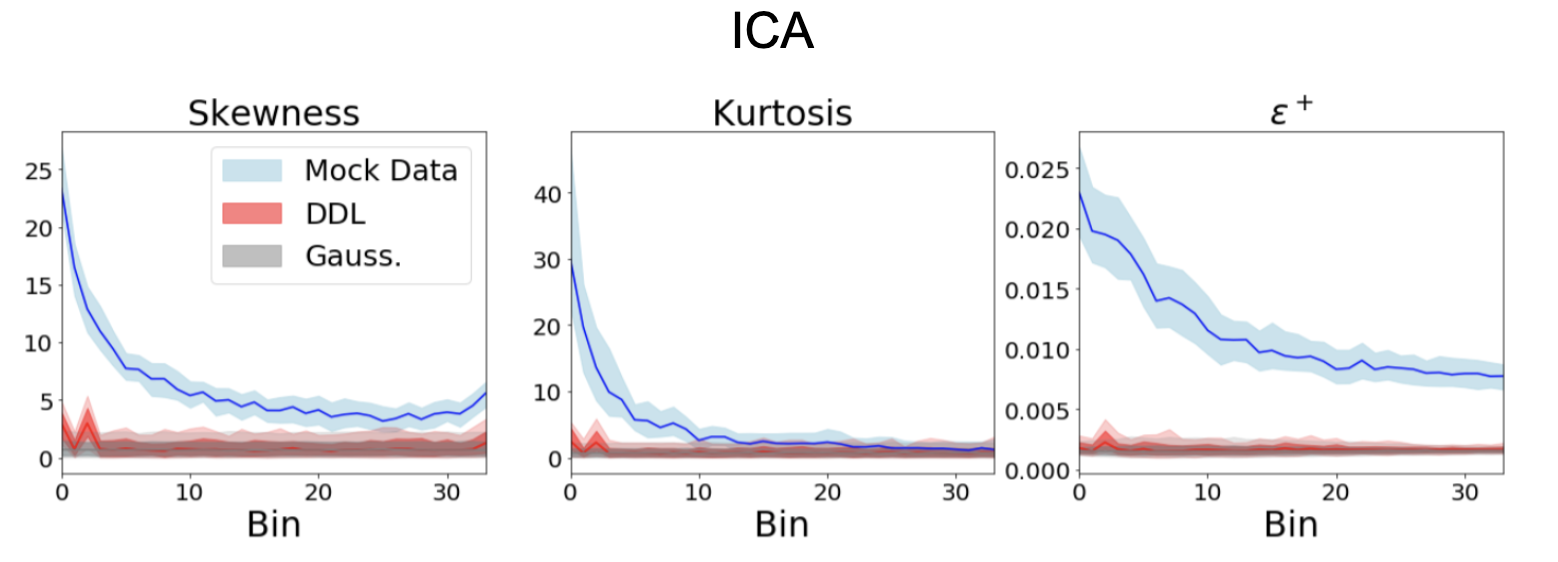}
    \caption{\footnotesize{ Same as the first three panels for the ICA likelihood in Figure \ref{fig:DDL_samples}, but fit on 2,048 mocks instead of the full 75,000. From left to right: absolute value of the $t$-statistic of the skewness, kurtosis, and $\epsilon^+$ of each bin for 100 different sets of 2,048 mock WL power spectra (blue), ICA samples (red), and Gaussian samples (gray). A vertical offset between the blue and gray contours reflects the NG in the data. Lack of overlap between the red and blue contours indicates that the ICA likelihood has not succeeded in capturing the NG. }}\label{fig:NG_ICA_2048}
\end{figure*}

Something that we were interested in was how the learning was inhibited (or not) by using an approximately singular vs. nonsingular covariance matrix. As given above, the squared exponential kernel using bin numbers as values for $x$ and $x'$ yields a singular covariance. By adding ``noise" to the diagonal (adding the identity matrix), we can turn it into a full rank matrix.

The results training on data drawn from a non-singular covariance matrix are shown in Figure \ref{fig:toy}. The two rows of panel (a) show the difference between the reconstructed FBL covariance (precision) matrix and the true one, as well as between the sample MVN covariance (precision) and the true one. It can be seen that the error in the FBL matrices is of the same order than that of the sampled MVN. 
The left figure in panel (b) shows the mean, 68\% and 95\% CLs from 2,000 samples drawn from a MVN with mean and covariance given by Eqs. (\ref{eq:mu}) and (\ref{eq:K}) in blue, and the mean, 68\% and 95\% CLs obtained from sampling the learned FBL in red. The right one shows the difference between the true and sample FBL mean (dashed red) and true and sample MVN mean (blue), which shows that the error in the FBL's mean is on the same order of magnitude as the sampling error.

Finally, panel (c) shows what the values of the log-likelihood are under a MVN for the test data with different combinations of $\mu$ and $\Sigma$, in blue. The green line common to all the subpanels is the true log-likelihood, and the red points are the log-likelihood given by the trained FBL for the same data. Clearly, the small scatter about the green line visible in the FBL values is of the same order than that due simply to sampling error when estimating the covariance matrix from a finite number of samples. 

Figure \ref{fig:singular_toy} shows the analogous results but for the case where no constant is added to the diagonal terms of the covariance matrix and it is thus singular. Everything else is the same as in the toy problem above. The interesting thing to notice is that, although the sample quality is still excellent, the quality of the likelihood is significantly worse. In panel (c) we see that the FBL has a much larger scatter as well as a constant offset with respect to the true likelihood. Furthermore, the MVN likelihood is much more sensitive to tiny deviations away form the true mean, as can be seen in the middle-bottom subpanel: even though the difference between the FBL sample mean and the true mean is very small, the likelihood values are abysmal. This is a general feature of MVN likelihoods with (nearly) singular covariances. Comparing the true and reconstructed precision matrices also reflects the fact that the model has not been able to learn the likelihood correctly. The issue of singularity is also discussed in Ref. \cite{bohm2020probabilistic}.

The dissonance between sample quality and likelihood quality is very interesting, and emphasizes the value in scrutinizing the likelihoods that flow-based models are learning, as we do in this work. The point is that simply sampling from the FBL and looking at the distribution of generated samples does not guarantee that the likelihood learned by the model is actually correct. 

\section{Data-driven likelihoods in data-limited regimes}\label{app:data_lim}

\renewcommand{\thefigure}{D\arabic{figure}}

\setcounter{figure}{0}

\begin{figure*}[t]
\centering
    \begin{subfigure}[b]{0.21\textwidth}
        \includegraphics[width=\textwidth]{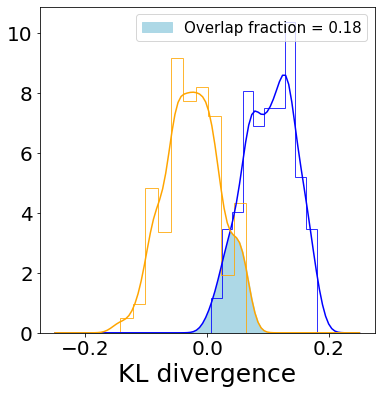}
    \end{subfigure}
    \begin{subfigure}[b]{0.205\textwidth}
        \includegraphics[width=\textwidth]{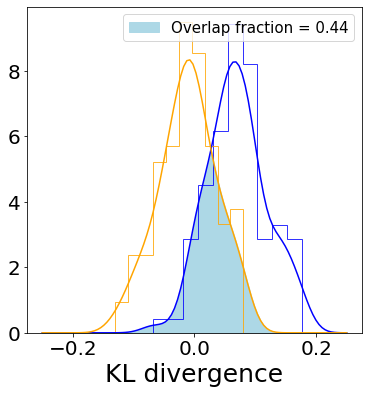}
    \end{subfigure}
    \begin{subfigure}[b]{0.21\textwidth}
        \includegraphics[width=\textwidth]{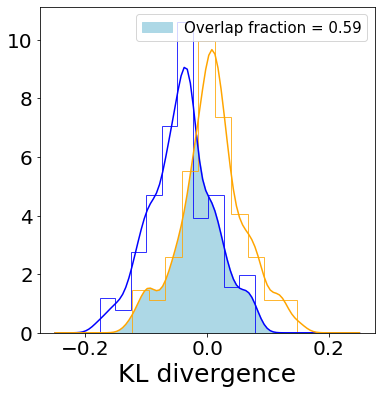}
    \end{subfigure}

\caption{\footnotesize{KL divergence estimate (Eq. \ref{eq:kl}) of two ensembles of draws with respect to the same reference distribution, all drawn from a 34-dimensional multivariate normal likelihood with mean and covariance from the weak lensing convergence power spectra mocks. Each panel corresponds to a different random seed. }}
\label{fig:kldiv_mvn}
\end{figure*}

As their name suggests, data-driven likelihoods can only truly thrive with plentiful data. Ultimately, these methods are trying to estimate densities (often in very high-dimensional spaces).
We illustrate the potential shortcomings of reaching conclusions by using DDLs fit on a limited number of mocks in Figure \ref{fig:NG_ICA_2048}. It shows the same results as in Figure \ref{fig:DDL_samples} for the ICA likelihood, but fit on 2,048 convergence power spectrum mocks instead of 75,000. Comparing these two figures it is clear that $2,048$ mocks are not enough for the DDL to capture the skewness and kurtosis in the first few bins. It is therefore possible that works that have operated in such data-limited regimes (e.g., \cite{Hahn:2018zja,cnn_wl1} have underestimated the effect that NGs can have on inferred parameters. 

\renewcommand{\thefigure}{E\arabic{figure}}

\setcounter{figure}{0}

\begin{figure*}
    \centering
    \begin{subfigure}[b]{0.8\textwidth}
        \includegraphics[width=\textwidth]{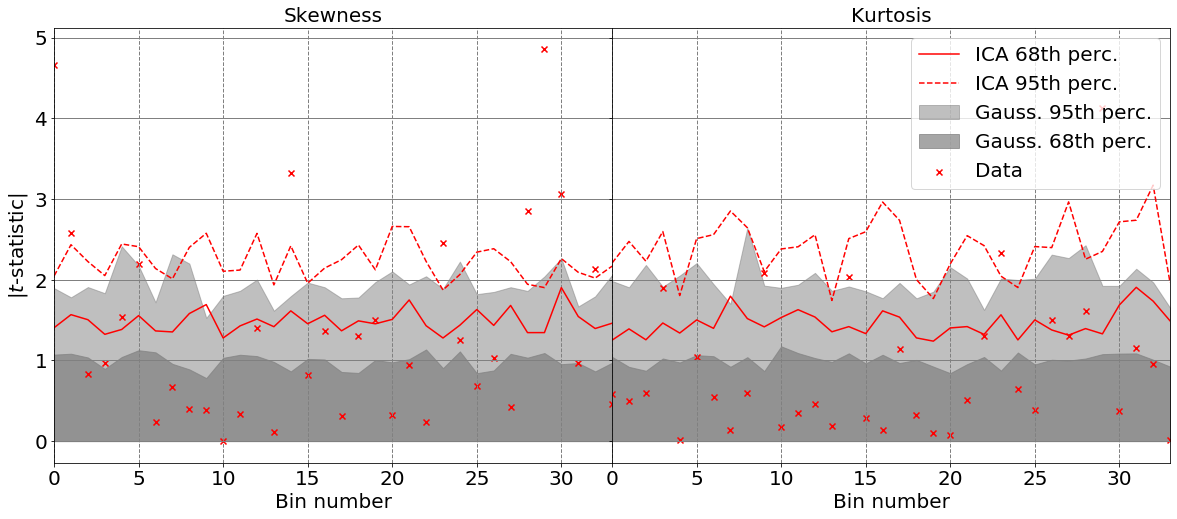}
    \end{subfigure}
    \begin{subfigure}[b]{0.8\textwidth}
        \includegraphics[width=\textwidth]{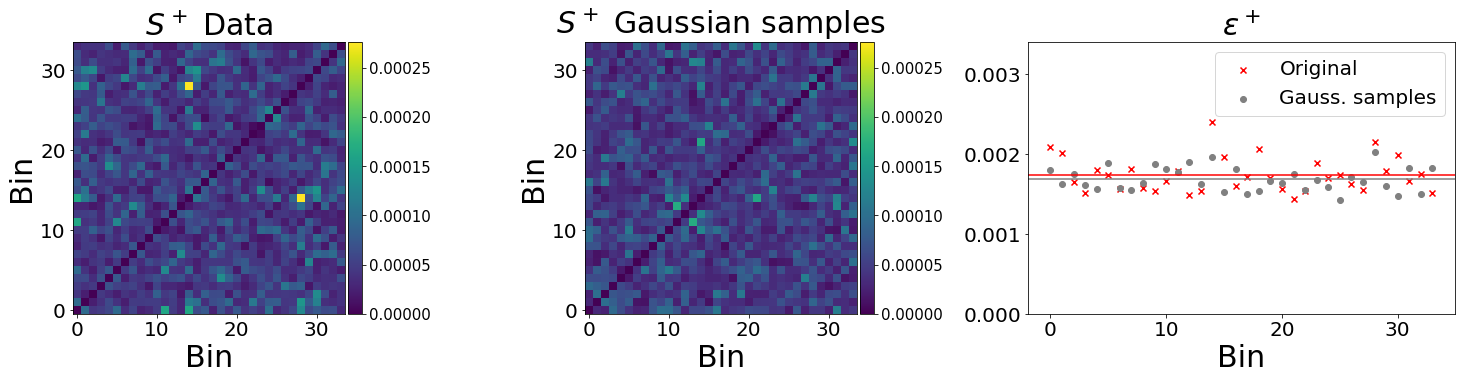}
    \end{subfigure}
    \begin{subfigure}[b]{0.25\textwidth}
        \includegraphics[width=\textwidth]{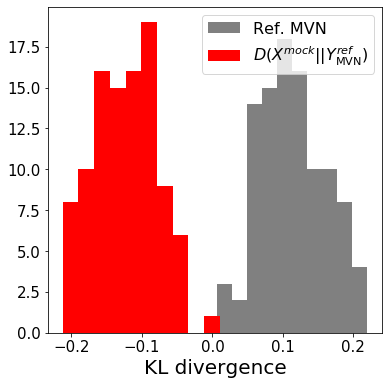}
    \end{subfigure}
    ~
\caption{\footnotesize{Same as Figure \ref{fig:WL_nongauss} but for the BOSS DR12 mock galaxy power spectra, as a function of bin number. \textit{Top}: the red crosses are the absolute value of the $t$-statistic of skewness (left) and kurtosis (right) of individual bins for 2,048 mock galaxy power spectra. The gray contours correspond to the one (dark gray) and two sigma (light gray) confidence level when averaging 100 different sets of 2,048 samples drawn from a multivariate normal likelihood with the same mean and covariance as the mock data. As expected, they correspond to $t$-statistic values of 1 (dark gray) and 2 (light gray). In addition, the red solid and dashed lines show the 68\% and 95\% CLs for 2,048 draws from the ICA likelihood fit on the data. Notice that ICA fails to capture the strong deviations from Gaussianity in the skewness. \textit{Middle}: the $S
^+$ matrix for the mock data (left) and equivalent Gaussian samples (middle). The sum over columns of each matrix, $\epsilon_u^+ \equiv \sum S^+_{u,v}$, is shown on the right as red crosses and gray circles, respectively. \textit{Bottom}: nonparametric KL divergence estimate between the mock data and their Gaussian counterparts (red), and the Gaussian samples with themselves (gray). The fact that the gray histogram is not perfectly centered at zero is due to the slight variability of the KL estimator in 34 dimensions , given the number of mocks considered (Appendix \ref{app:kldiv}).}}\label{fig:tstats_crosses_BOSS}
\end{figure*}

\section{Robustness of the nonparametric KL divergence estimator} \label{app:kldiv}

In this appendix we test the robustness of the nonparametric KL Divergence estimator introduced in Section \ref{sec:nongauss}. To do so, we estimate the same reference KL divergence histogram as the one shown in gray in Figure \ref{fig:tstats_crosses_BOSS} for two different Gaussian datasets drawn from the same likelihood. Figure \ref{fig:kldiv_mvn} shows the results using different random seeds. It can be seen that the degree of overlap can vary quite a bit, even from samples drawn from the same underlying distribution. The leftmost panel, with an overlapping area equal to 0.19, is the lowest we found. Therefore, while the large horizontal offset seen between the reference and data histograms in Figure \ref{fig:WL_nongauss} and Figure \ref{fig:DDL_samples} still seems statistically significant, the one seen for GMM2 in Figure \ref{fig:DDL_samples} does not. This is likely due to the curse of dimensionality, since the KL estimator relies on a $k$NN algorithm, and further emphasizes the importance of including more fine-grained measures of NG that are more robust with limited numbers of samples when quantifying non-Gaussianity.

\section{Galaxy power spectrum non-Gaussianity}\label{app:BOSS}

Ref. \cite{Hahn:2018zja} studied non-Gaussianities in two large-scale structure observables, one of them being the galaxy power spectrum. They used the MultiDark-PATCHY mock catalogs \cite{multidark}, which were built to match the BOSS Data Release 12 observations. In particular, they looked at the 2,048 mocks for the North Galactic Cap (NGC) in the redshift bin $0.2 < z < 0.5$ and obtained mock power spectra using \texttt{NBodyKit} \cite{nbodykit}. They used the nonparametric KL divergence test described in Section \ref{sec:nongauss} to establish the likelihood non-Gaussianity. 

\begin{figure*}
    \centering
    \includegraphics[width=0.9\textwidth]{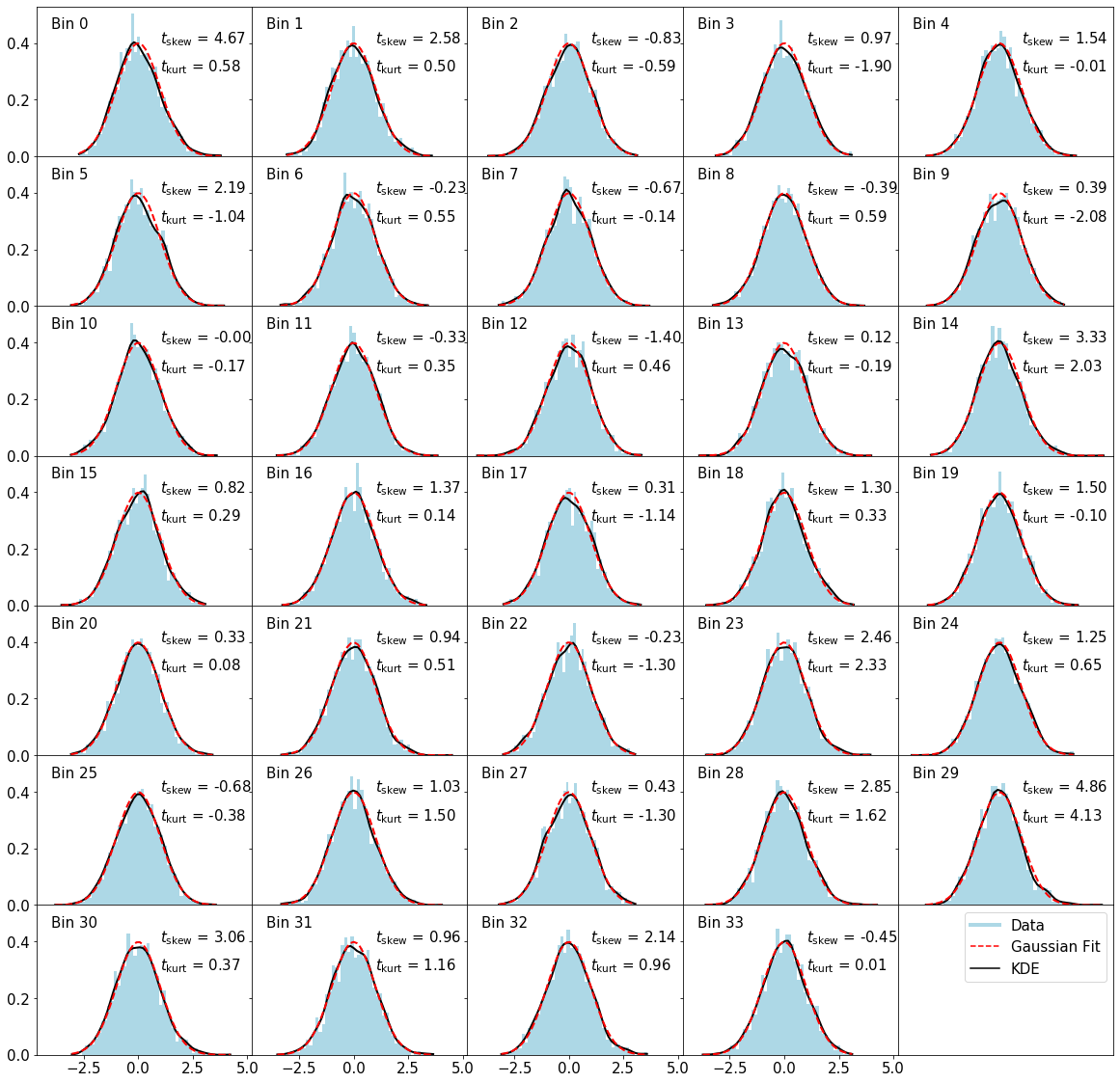}
    \caption{\footnotesize{Distribution for each of the 34 bins in the ensemble of 2,048 mock galaxy power spectra for the BOSS DR12 North Galactic Cap in the redshift bin $0.2 < z < 0.5$. The histogrammed values are shown in light blue, while the KDE is shown in black. The dashed red line is a Gaussian fit to each bin.   }}\label{fig:tstats_panels_BOSS}
\end{figure*}

They then sought to build a data-driven likelihood that would incorporate the NG. They showed that the estimated KL divergence was unchanged with a GMM likelihood\footnote{When we try fitting a GMM to the same mock dataset, we find that the BIC is minimized for a single component. The fact that the BIC increases monotonically with more components is due to the number of mocks being too small to fit the large number of parameters without overfitting. For example, with our WL mocks we find that with 2,048 samples the BIC behaves similarly, while with more mocks it finds the minimum at $K=2$.}, but nearly vanished with an ICA likelihood. They used this likelihood to perform importance sampling on an MCMC chain that had been previously analyzed with a Gaussian likelihood \cite{beutler} (essentially re-weighting the points by the ratio of the ICA likelihood to the MVN likelihood) and found small shifts ($<0.5\sigma$) in relevant cosmological parameters. 

We produced mock galaxy power spectra from the MultiDark-PATCHY catalogs using the same procedure as Ref. \cite{Hahn:2018zja}, so we refer the reader to them for details. Ultimately we obtain 2,048 mock power spectra in 37 bins, encompassing the power spectrum monopole, quadrupole, and hexadecapole. For the remainder of this section, however, we truncate the power spectrum at 34 bins, so that the NG tests (Section \ref{sec:nongauss}) are directly comparable to the ones in Section \ref{sec:wl_dset} for the WL power spectra (recall that the tests are extensive). Note that this means that our measures of non-Gaussianity are conservative compared to those in Ref. \cite{Hahn:2018zja}. Just like we did for the WL mocks, we mean-subtract and whiten the data before running it through the NG tests.

Figure \ref{fig:tstats_crosses_BOSS} is analogous to the one shown for the WL mock data in the main text showing the non-Gaussianity results from our three tests. In addition, Figure \ref{fig:tstats_panels_BOSS} shows the individual galaxy power spectrum bin distributions. Comparing the values of the $t$-statistics here and the ones for the WL mocks reveals that the skewness and kurtosis are much more pronounced in the latter. The same can be said when comparing the mock and Gaussian $S^+$ matrices, as well as the vertical offset between the mock and Gaussian $\epsilon
^+$. 

Keeping in mind the discussion in the main text on the applicability of different DDLs depending on the type of NG in the data, we can see that the fact that $\epsilon^+$ is virtually indistinguishable between the data and the Gaussian samples, while some individual bins do exhibit non-negligible skewness and kurtosis, could explain why the ICA likelihood worked better in this setting than the GMM did (though the small number of mocks might also be a contributing factor). 

The top two panels of Figure \ref{fig:tstats_crosses_BOSS} also show the 68\% and 95\% CLs from 2,048 samples drawn from the ICA likelihood. It can be seen that, while slightly higher than the Gaussian ones, few of the data points with strong NG fall within the 2$\sigma$ boundary of the ICA samples. This corroborates our results on WL data and Appendix \ref{app:kldiv}, which showed that even if the KL divergence is small between a catalog of data and a data-driven likelihood, the DDL may not be accurately capturing the NG in the data. 
We thus conclude that it is likely that the non-Gaussian signatures in galaxy power spectrum data could actually have a larger impact than was found in Ref. \cite{Hahn:2018zja}, but the limited number of mocks makes this a difficult task for a DDL to solve adequately.

\section{Mock weak lensing convergence maps}\label{app:WL_maps}

\renewcommand{\thefigure}{F\arabic{figure}}

\setcounter{figure}{0}

\subsection{Generating the mocks}\label{subsec:generating}

Generating mock WL maps involves running full $N$-body simulations, which consist of evolving millions of particles under the force of gravity to simulate the formation of structure from an early time, when the matter field was nearly homogeneous, to late times, where it is highly clustered. As a first step, we generated a primordial power spectrum for a given choice of cosmological parameters\footnote{We set $\Omega_{m}=0.3$, $\Omega_{\Lambda} = 0.7$, $\Omega_{\rm b} = 0.046$, $\sigma_8 = 0.8$, $n_{s} = 1$ and $H_0 = 72$ km/s/Mpc.} with the Boltzmann solver \texttt{CAMB} \cite{camb}. From the power spectrum we used \texttt{N-GenIC} \cite{ngenic} to generate the initial conditions for the particles in the simulation box, and finally \texttt{Gadget2} \cite{gadget2} to evolve the particles. We used a box with a comoving length of 240 Mpc/$h$ on each side and $512^3$ particles. The particles were initialized at redshift $z=100$, evolved until $z=0$, and snapshots were saved at 60 different redshifts between $z=3$ and $z=0$. For reasons described below, we ran four different $N$-body simulations, all with the same underlying cosmological parameters but with different seeds for the initial density and velocity perturbations. 

To generate the convergence maps we used the software package \texttt{LensTools} \cite{lenstools}, which implements a multi-lens-plane algorithm for ray-tracing. This algorithm approximates the three-dimensional distribution of matter $\delta(\boldsymbol{x},z)$ (obtained from an $N$-body simulation) between the source redshift $z_{\rm s}$ and us as a series of discrete two-dimensional planes perpendicular to the line of sight, with thickness $\bigtriangleup$ and surface mass density $\sigma$:

\begin{equation}
\sigma(\boldsymbol{x},z) = \frac{3 H_0^2 \Omega_{m}\chi(z)}{2 c^2 a(z)}\int_{\bigtriangleup}d\chi'\delta(\boldsymbol{x},z(\chi')),
\end{equation}

\noindent where $\chi$ is the comoving distance, $a = 1/(1+z)$ is the scale factor, $\Omega_m$ the matter density, $H_0$ the Hubble constant and $c$ the speed of light. Since the surface density and the gravitational potential $\phi$ are related via the two-dimensional Poisson equation,

\begin{equation}
\nabla_{\boldsymbol{x}}^2\phi(\boldsymbol{x},z) = 2\sigma(\boldsymbol{x},z),  
\end{equation}

\noindent the gravitational potential can be solved for. Then, the angular photon trajectory $\boldsymbol{\beta}$ can be calculated using the geodesic equation. Finally, the shear $\boldsymbol{\gamma}$ and the convergence $\kappa$ can be obtained, since they are elements in the Jacobian of the angular trajectory of a photon as a function of its initial position $\boldsymbol{\theta}$:

\begin{equation}
\frac{\partial \boldsymbol{\beta}}{\partial \boldsymbol{\theta}} = \begin{pmatrix} 
1-\kappa + \boldsymbol{\gamma}_1 & -\boldsymbol{\gamma}_2 \\
-\boldsymbol{\gamma}_2 & 1-\kappa - \boldsymbol{\gamma}_1 
\end{pmatrix}.
\end{equation}

To make the two-dimensional density planes we cut each snapshot at three points (55, 167, 278) Mpc and project slabs (along all three axes) that are 111 Mpc thick around each cut point. The planes are generated at a resolution of $4096 \times 4096$ pixels to make sure that small-scale information is preserved. To build the convergence maps we place a source at $z_{\rm s} = 1$ and 30 planes between $z_{\rm s}$ and $z=0$, where a plane at a given redshift is randomly chosen from the planes made from each of the four different $N$-body simulations at that redshift. This is done to generate random realizations of the convergence field that are statistically independent. We created $1024 \times 1024$ pixel convergence maps that have a sky coverage of $3.5 \times 3.5$ deg. The left panel of Figure \ref{fig:mock_data} shows an example of a simulated convergence map. We obtained the convergence power spectra through \texttt{LensTools} as well. We ultimately obtain 75,000 mock power spectra in 34 bins, uniformly distributed in log space for $\ell = [100,5000]$, where the lower limit is set by the size of the map and the upper limit by the fact that at higher multipoles the numerical power spectra were shown to deviate from theory significantly \cite{cnn_wl1}. The 68\% CL and 95\% CLs of our mock observable can be seen in the right panel of Figure \ref{fig:mock_data}.

\subsection{Convergence power spectrum distribution per bin} \label{app:tstats_panels_WL}

\begin{figure*}
    \centering
    \includegraphics[width=0.9\textwidth]{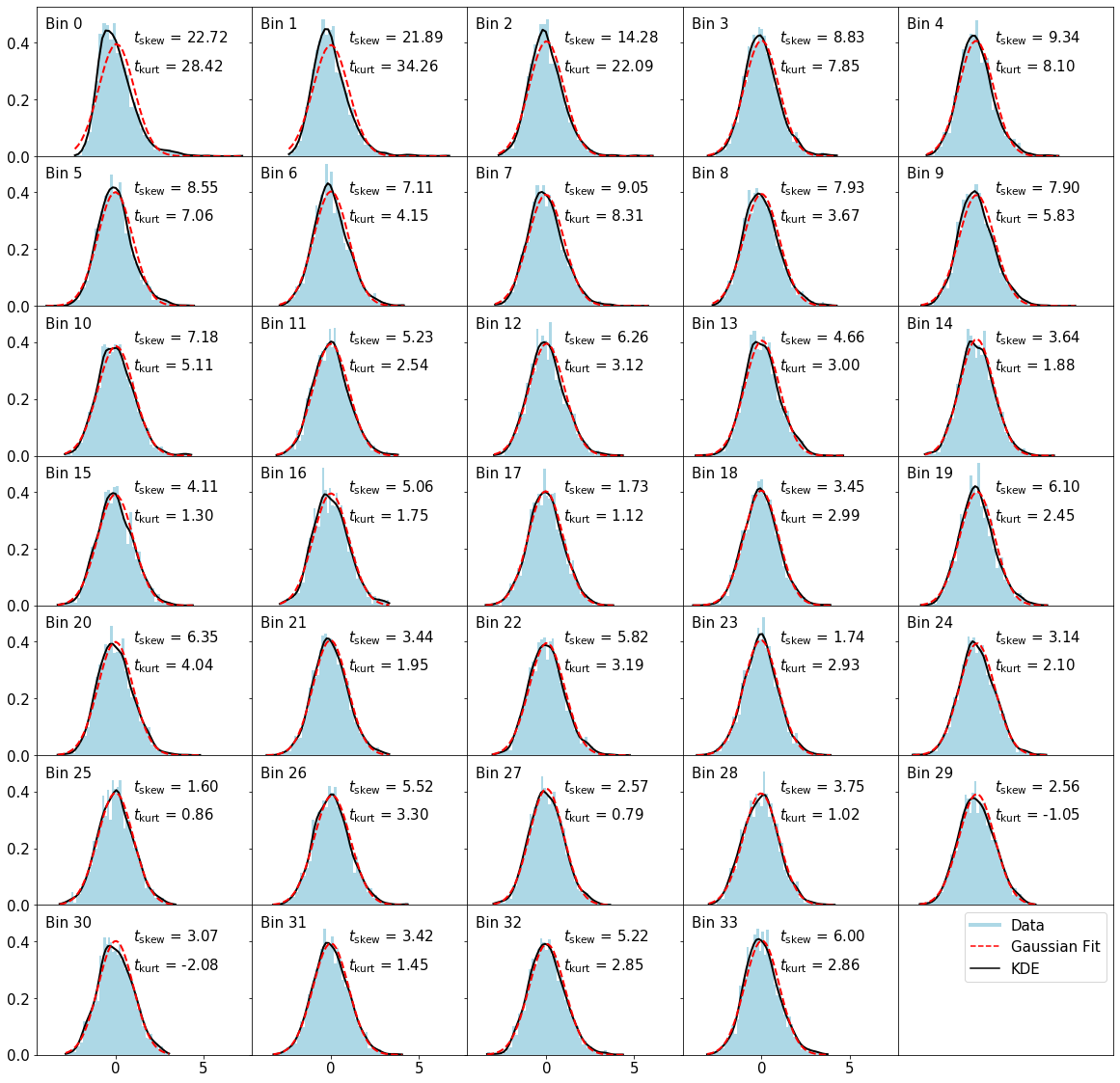}
    \caption{\footnotesize{Distribution for each of the 34 bins in an ensemble of 2,048 mock convergence power spectra. The histogrammed values are shown in light blue, while the KDE is shown in black. The dashed red line is a Gaussian fit to each bin. }}\label{fig:tstats_panels_WL}
\end{figure*}

In Section \ref{sec:wl_dset} we showed the absolute value of the $t$-statistic for each bin in the ensemble of 2,048 mock convergence power spectra. In Figure \ref{fig:tstats_panels_WL} we show the actual distribution for each bin as blue histograms. We also show a Gaussian fit to the histogram in dashed red, and a KDE of the distribution in black. For many of the bins, the deviation away from zero skewness and kurtosis can be seen by eye by comparing the Gaussian fit to either the KDE or the histogram.

\bibliography{main}

\end{document}